\documentclass[prb,twocolumn]{revtex4}
\usepackage[dvips]{graphicx} 
\usepackage{epsfig} 
\usepackage{epstopdf}
\usepackage{axodraw}
\usepackage{amssymb}

\newcommand{\bea}{\begin{eqnarray}}
\newcommand{\eea}{\end{eqnarray}}
\newcommand{\beq}{\begin{equation}}
\newcommand{\eeq}{\end{equation}}
\newcommand{\benu}{\begin{enumerate}}
\newcommand{\enu}{\end{enumerate}}

\begin{document}
\title{Selective Mott  transition and  heavy fermions }
\date{\today}
\author{C. P\'epin}
\affiliation{SPhT, CEA-Saclay, L'Orme des Merisiers, 91191
Gif-sur-Yvette, France\\
}
\begin{abstract}
    Starting with  an extended version of the Anderson lattice where the f-electrons are allowed a  
weak dispersion, we examine the possibility of a Mott localization of the f-electrons,  for a finite value 
of the hybridization $V$.   We study the fluctuations at the quantum critical point  (QCP) where the f-electrons localize. 
We find they are in the same universality class as  for the Kondo breakdown QCP, with the following notable features. 
 The quantum critical regime sees the appearance of  an additional energy scale separating two universality classes. 
In the low energy regime, the fluctuations are dominated by massless gauge modes, while in the intermediate 
energy regime, the fluctuations of the modulus of the order parameter are the most relevant ones. 
In the latter regime, electric transport simplifies drastically, leading to a quasi-linear resistivity
 in 3D and anomalous exponents lower than  T in 2 D. This rather unique feature of the quantum critical 
regime enables us to make experimentally testable predictions. 
\end{abstract}

\leavevmode
\maketitle

\section{Introduction}

  Several years  of intense  experimental studies  of quantum criticality in inter-metallic 
and heavy fermion compounds has lead to  a growing  evidence for a violation of the standard 
Landau Fermi liquid theory of  the metallic behavior. 
 Experimental phase diagrams use an external
 tuning parameter, like the chemical pressure, the hydrostatic pressure or  magnetic field to
 explore phase transitions at a temperature very close to the absolute zero.   These phase 
transitions exhibit a regime of very strong  quantum fluctuations -the quantum critical regime-,
 where  anomalous  thermodynamic and transport exponents are observed.  A heavy fermion compound  
consists in a lattice of  ``big'' atoms, like Ce , U or Yb, alloyed with a metal.  The magnetic
 multiplet in the heavy atoms is determined by Hunds rules,  and  spin orbit interaction. 
 For example, after spin-orbit interaction, the magnetic moment of Ce -4f$^1$  ($S=1/2$, $L=3$ ) is  
$J=5/2$, for  Yb-4 f $^{13}$ ($S=1/2$, $L=3$)  we get $J=7/2$  and for  U 5f$^2$ ($S=1$, $L= 3+2$) it
 leads to  $J=4$. The degeneracy is then split by the lattice crystal field effects, finally leading to
 a Kramers doublet.   The  apparent chemical complexity of these material has to be kept in mind for any 
investigation of  their anomalous experimental properties. In particular, the Anderson lattice model, 
which is  standard in describing those  compounds, relies upon the assumption that the Kramers doublets 
are well formed\cite{note1} and thus, at low energy, 
  one  effectively considers that the physics is 
  thoroughly  described  by f-electrons, sitting on the impurity atoms,  subjected to  strong 
  Coulomb interactions , and hybridizing with the conduction electrons of the metal. 
  The beauty of the experimental  results lies in the fact that, although the compounds are based 
on a complex chemistry,  very clean anomalous universal exponents in both transport and thermodynamics
 are observed. In this paper, we don't detail the experimental results, but rather
refer the reader to the various reviews on the 
subject\cite{stewart, review-piers,lohneysen,qimiaosteglich}.
 One of the most famous example of universality is the linear in $T$ resistivity observed in YbRh$_2$Si$_2$ 
 fine tuned to quantum criticality with a small magnetic field parallel to the c-axis.  In this experiment , 
the linearity of the resistivity persists on three decades of energy, which makes it one of the most robust 
 anomalous exponents  in strongly correlated materials\cite{note2}. The coefficient of the specific heat
 $\gamma = C/T$ is found to increase when getting closer to  the QCP without showing any sign of saturation. 
A logarithmic law is generally attributed to this increase, but in several case, like for YbRh$_2$Si$_2$  
a divergence stronger than logarithmic is  inferred from the data\cite{review-piers}. 
  Recently another type of experiment showing quantum criticality has appeared with  the observation of strong 
 effective mass increase in He$^3$  bi-layers \cite{saunders}.  In this experiments,   one 
layer of  highly compressed solid He$^4$ is adsorbed on a graphite substrate, then another layer 
of  solid He$^4$,  and on top of it, two layers of He$^3$ are adsorbed. The first layer of  He$^3$ 
undergoes a transition towards a localized state at a finite filling factor. Measurements of the specific
 heat  show a strong increase of the effective mass  inversely proportional to the coverage,
 associated with a decrease of the coherence temperature  in $ (n-n_c)^{1.8}$,  as the QCP is reached. 
  \begin{figure}[h]
\includegraphics[width=3 in]{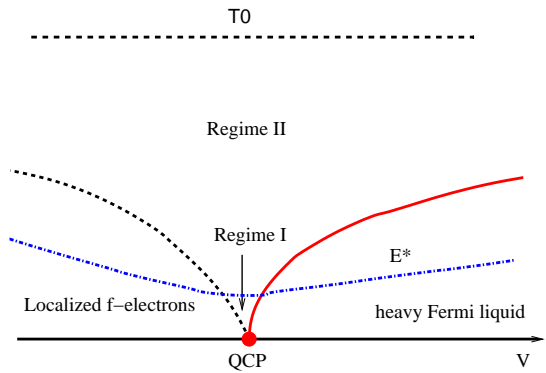}
\caption{ Schematic phase diagram for the  selective Mott of the f-electrons in the Anderson lattice. 
On the right, when the holons condense, is the heavy Fermi liquid phase, also called in this paper the Higgs phase. 
On the left, where the  holons are not condensed, is the localized phase for the f-electrons. We don't analyze here
 the magnetic character of this phase but rather characterize it only by the fact that the f-electrons localize.
 The QCP is multi-scale as  is shown by the scale $E^*$. Two regimes are of interest: (i) for $T\leq E^*$ the 
 exponents depend on the shape of the Fermi surface, while (ii), for $ T \geq E^*$  the exponents are universal 
  and  show quasi-linear resistivity in $ D=3$.  $T_0$ is the upper energy cut-off, above which the entropy   $ R ln 2 $ 
  per site is released.} \label{phasediagram}
\end{figure} 

     It is fair to say that this increasing body of fascinating experimental results is, at
 the present time, still mysterious.
     The biggest challenge for the theoretician is to account for the linear resistivity in 
three decades of energy, in 3 dimensions of space. Such a result calls for a specific scattering 
process, from which the lifetime of the conduction electron would acquire the characteristic  linear
 in T dependence.  At the moment, no theory can produce a reasonable scenario accounting for  
linear-in-T resistivity down to zero temperature, in three dimensions of space.  The present paper, 
 however, produces a very simple explanation for  a resistivity linear in T , but in the 
intermediate  quantum critical regime; namely above a finite (although quite small) energy scale.

      Since the beginning of the  history of heavy fermions, physicists have been
 struck by the intense magnetic nature of these materials. The presence of big atoms and
 big moments naively calls for magnetism. Moreover,  magnetic interactions mediated 
 by conduction electrons - called RKKY interactions - are present in these materials\cite{RKKY}. They 
compete with the formation of the Kondo singlets responsible for the condensation of the 
heavy Fermi liquid phase\cite{levin,millis-lee}.  For years the magnetic nature of the materials was so 
overwhelming that it  was a real surprise when the first heavy fermion superconductor 
was discovered\cite{steglich}. Similarly, the 
      competition between magnetism and  the formation of the Kondo singlets has been 
 considered until recently to be the main physical forces at play in the Anderson lattice \cite{doniach,lacroix}.
      In a completely natural way, the theory has first focused on  QCPs separating a 
metallic magnetic phase from a metallic paramagnet\cite{hertz,millis,moriya}. This is this the simplest case of a 
QCP for itinerant electrons, usually called spin density wave (SDW) scenario.  Here the  
Fermi surface of the conduction electrons is destabilized by scattering through magnetic
 modes propagating through the metal: the paramagnons. A key concept in this theory is 
the dynamical exponent $z$. Since close to a QCP quantum fluctuations are 
always relevant, the correlation  in  imaginary time has to be taken into account which
 leads to  an effective  dimension $d + z$ larger than the  $d$-dimension of space.  
In practice, $z$ is given by the structure of the bare paramagnon propagator.   In the 
case of a transition towards an anti-ferromagnet (AF),  the  paramagnon propagator has 
the form $D^{-1}( q, i \omega_n ) =  c  |\omega_n | + q ^2 $ where $ c $ is a constant. 
Hence $ z=2 $ here.    The theory  then integrates  the fermions out of the partition 
function\cite{hertz,millis} -step which is not rigorously justified -, to obtain  an effective $\phi^4$ 
lagrangian in $d+z$ dimensions. Since $z=2$  in the AF case, one is  typically  above 
the upper critical dimension of the model and the bosonic effective theory can be solved 
at the mean field level. A better treatment of the model doesn't integrate  the 
 fermions out of the partition function, but  relies on an Eliashberg-type theory  where 
the vertices are neglected  and self-energies conserved\cite{chubAF,jerome}.  This theory  is 
controlled in  a rather artificial  large N approximation where N multiplies  the number
 of fermions. However, we use it in this paper every time we deal with a QCP of itinerant 
electrons, since it is the best technique available at the moment in that it  avoids  
integrating the fermions out  of the partition function, hence not missing any infra-red (IR) divergence.
       Intense theoretical studies of the SDW scenario lead to the conclusion that  some
 heavy fermion might fall into the universality class of the 3D AF, like CeNi$_2$Ge$_2$\cite{rosch}.  
Paramagnons, however, cannot account for strongly anomalous exponents like a linear resistivity in 3 D. 
       
         The search for new QCPs then started. New ideas emerged, like the one 
of ``de-confinement''  of  the heavy quasi-particle, close to  the magnetic phase transition\cite{review-piers}.
  Meanwhile, a phenomenological approach was proposed,  based on the observation that 
one can fit the NMR data for many compounds with two kinds of elementary excitations, 
that the authors called ``two fluids''\cite{pines}. Although it's not clear whether it's really two 
``fluids'' that are  the good elementary excitations, the phenomenology 
looks robust and puts a landmark in the landscape of the field: the right microscopic theory
  should  eventually reproduce the two fluids phenomenology.
        The first idea that a local mode appears at the QCP was given in \cite{qimiao} 
in the so-called ``locally quantum critical scenario''. The strength of this idea is to  
have been the first to outline that ``something local'' has to appear at the QCP.  In this
 theory,  the local mode is of magnetic nature (a local spin). Its weakness, however, is that it requires 
the magnetic structure of the material to be purely  two-dimensional\cite{pankov}.  Moreover it appears 
to be very difficult  to put it to the experimental test, because the technique  involved  is not obviously 
transparent.  To date, no experimental prediction has been made which would allow the community 
to accept, or discard it.

          Then, came the idea of  a reconstruction of the Fermi surface at the quantum critical
 point, going with the idea of ``de-confinement''\cite{review-piers}.   The first tractable effective field 
theory illustrating the ideas of  de-confinement and of Fermi surface reconstruction was 
given in \cite{ssv}. The authors used  the Kondo- Heisenberg  (K-H) model, and
 showed that, at some point in the phase diagram, there is a  transition towards a spin liquid
 phase, that they called FL$^*$. At the transition, the Fermi surface re-constructs and one 
looses half of the charge carriers in the phase FL$^*$. Another way to look at the transition,
 is to say that, in the heavy Fermi liquid phase,  the f-electrons start to conduct, as 
was beautifully shown by \cite{schofield}. Apparently the theory seemed to stand on much more solid 
grounds, since for the first time there was a tractable model on  which to work. However, the 
treatment \cite{ssv} suffered some insufficiencies.
          The most dramatic one is the complete lack of  plausible experimental predictions.
 Apparently, the authors were not able to reproduce the linear in T resistivity, and the
 exponents they found for the anomalies  in the specific heat coefficient didn't match the experiments.
          Moreover there was the problem of the phase FL$^*$. To have a ``spin liquid'' phase 
in any heavy fermion compound appeared to be very unlikely.  The most  promising candidate 
for it, was the system of He$^3$ bi-layers, for which intensive numerical studies in the 
last decade\cite{roger,misguich} have shown with very little doubt that,  in its localized phase,  adsorbed He$^3$ layers form a 
    spin liquid.

    \section{Relation to previous work and structure of the paper} 
          
           This paper details  the theory of the ``Kondo breakdown'', previously
     exposed in two letters\cite{us,cath}. We describe the excitations around a selective Mott transition in 
the Anderson lattice, using the simplest theoretical tool available, the U(1) slave-boson 
gauge theory.  Our goal is to  cast the theory into the  Eliashberg formalism 
 with a careful treatment of the gauge invariance.  To our knowledge, it is the first time 
 that the Eliashberg technique is presented  for  a ``de-confined'' QCP, where the order parameter 
 breaks a gauge symmetry (instead of a global symmetry ).  In particular, we present a study of 
 the transport around 
 such a  QCP, using the Ioffe-Larkin (IL) composition rule. We derive as well  the
  most generic Ward Identities (WI) associated with the 
 gauge symmetry.
 
  First note that the first study of a phase transition separating a spin liquid 
  from a heavy Fermi  liquid phase was performed
 for  the Kondo 
 lattice model with frustrated magnetism, using the 
 DMFT technique \cite{burdin}.
 
  Two other studies exist. 
 The authors of \cite{schofield} have shown that, at such a QCP in the K-H model, there is a jump 
 in the transport properties; the residual resistivity jumps at the QCP and the Hall conductivity as well. 
 This follows from the fact that the f-electrons start abruptly to conduct in the heavy Fermi liquid phase.
These results come naturally in our formalism, we thus agree with  \cite{schofield}. Our paper 
 re-derives their results using the
simpler formalism of the IL composition rule.

 The second study is \cite{ssv}. The authors  were the first to show that the volume of the Fermi surface jumps at the QCP.
 We agree with this result and with their observation that the thermodynamics,
 in the vicinity of the QCP, is dominated by the gauge fields.
We disagree with their findings about transport.  The authors of Ref.\cite{ssv} claim that at very low energy transport is dominated by the
lifetime of the holons. One message of our study, is that transport is dominated  by the conduction electrons in all regimes\cite{note10}. This is imposed by the IL composition rule
\beq
\rho= \rho_c + \left ( \rho_f^{-1} +\rho_b^{-1} \right ) ^{-1} \ , \eeq which states that the resistance of the holons is in series with  the resistance  of the spinons  and in parallel with the resistance of the conduction electrons.
 
 Moreover, as already mentioned in \cite{us,cath},  the general structure of the QCP is much richer than what was found in
 \cite{ssv}. This is depicted in Fig\ref{phasediagram}.  
  The first overlooked point is that the QCP 
 is multi-scale. By this, we mean that  an  intermediate energy scale is present 
at the QCP, separating two  quantum critical regimes of universal exponents. The second 
overlooked point is that, in the intermediate energy regime, electrical conductivity simplifies,
 and one obtains a resistivity quasi-linear in T, in 3 D.  This regime
 is of main importance, because it enables us to connect to experimental results and make 
predictions to test the theory.   A first application of  this regime to  He$^3$ bi-layers 
has already been given\cite{adel}.  The third overlooked point is that, at the mean field level, a modulated 
solution of the heavy  Fermi liquid state exists, analogous in structure to the modulations 
found for FFLO superconductivity\cite{FFLO}. 
An analogous state was found in the study of Chromium \cite{rice}. 
This state may  be of relevance for  the mysterious 
phase observed in CeCoIn$_5$ where the magnetic field is applied  in the (ab)  axis.
           Last, it appears that,  in the case of the more physical model of the Anderson 
lattice with a small dispersion of the f-electrons, the Kondo breadkdown  QCP coincides 
with a Mott transition of the f-electrons\cite{cath}.   

The main advantage of this viewpoint is to
 simplify the discussion and to connect with other techniques available,
  like Dynamical Mean Field Theory.  DMFT studies can now be performed which will confirm or infirm
 in the short term the mean field findings of a selective Mott transition in the periodic Anderson lattice model (PAM). 
 Selective Mott transition has been previously studied in the context of multi-orbital Hubbard models
\cite{knecht,ferrero,bunemann} where transitions for various bands have been found.
This model differs from the PAM because of the the absence of hybridization between the bands.
 Recent DMFT studies\cite{demedici,deleo,ping,sordi} of the PAM
 tend to say that the selective Mott transition indeed exist, but the  
 studies are not completely
 conclusive yet.

  The second advantage is  to link the 
discussion with the physics of high temperature superconductors\cite{anderson,affleck-marston,lee-review,leenagaosa}. 
 It is notorious that the 
conduction electrons undergo a Mott transition in the phase diagram of high  T$_c$.  Take a model with no frustration, 
like the K-H model on the square lattice. One can show 
 formally that, in the localized phase, the model is equivalent to a Heisenberg anti-ferromagnet
and that its ground state is an anti-ferromagnet. This point was actually made in Ref\cite{ssv}.
Although this statement is formally simple, it is very difficult to obtain within the slave boson technique.
 In 2 D, for high T$_c$ 
superconductors, one can advocate that the spin liquid naturally re-confines under the effect of  
gauge fluctuations\cite{leenagaosa}. We don't have this latitude in 3 D where gauge fluctuations are benign and
 can be treated  within a non compact formalism.   A better route towards the AF ground state 
is probably to allow  the system  to have both AF order and spin liquid, and then study the issue
 of re-confinement in that case.
       Similar studies  have been performed in the early days of high T$_c$\cite{hsu} but have
 not yet been done for the Anderson lattice.

         On the theory side, the Kondo breakdown QCP suffers as well from its own weaknesses.  
We  outline the main one, in  our view, which is that this fixed point relies on the presence 
of a spin liquid  at the transition.  This spin liquid is described in terms of  massless spinons, 
and the model is very sensitive to the Fermi surface of the spinons. To be precise,  the exponents 
obtained using, for example, the uniform spin liquid are different from the ones obtained using a nodal spin liquid.
 Whether the linear in T resistivity found in  the intermediate phase survives the ``spinonology''
 is still an open question.  A related question is to ask whether the U(1) gauge theory used 
in the description of the Kondo breakdown is the adequate tool to describe the approach to a Mott 
transition.  Are the elementary excitations  correctly captured within this technique ? 
         This question is  of broad interest and is as well at the heart of the physics of  high 
temperature superconductors.
         
           The paper is organized as follows. In section \ref{sec:model}
we recall the model and introduce the slave-boson effective lagrangian. Section \ref{sec:MF} and \ref{sec:modulations} are devoted to the mean-field 
approximation.
    Precisely, in section \ref{sec:MF} we  describe the mean-field approximation, showing the presence of 
 the QCP at  $T = 0$.  In section \ref{sec:modulations} we recall the possibility of another mean-field solution:  a modulated 
 order parameter in the heavy Fermi liquid phase.  
 In sections \ref{sec:amplitude-fluc} to \ref{sec:eliashberg} we study the fluctuations. We first start, in section \ref{sec:amplitude-fluc} and \ref{sec:gauge-fluc},  by introducing the amplitude and gauge fluctuations within a ``naive'' RPA approximation.  In section \ref{sec:eliashberg} we  expose the Eliashberg formalism which links self-consistenly the amplitude and gauge fluctuations, and is the best  available tool to study QCPs of that type. 
 Precisely, in section \ref{sec:amplitude-fluc}
  we study the fluctuations 
 of the amplitude  of the order parameter, showing the different regimes of quantum criticality. 
 In section \ref{sec:gauge-fluc} we introduce the gauge fluctuations, give 
 the form  of their propagator,  and detail the gauging out of the theory.    
  In  section \ref{sec:eliashberg} we recall the principles of the  Eliashberg theory and re-cast our study of  the 
  fluctuations in this framework. We then turn to electrical transport. In section \ref{sec:IL}, we compute the Ioffe-Larkin (IL) composition rule
   for the  conductivity around the QCP and derive the various transport  lifetimes in the low temperature regime.
  In section \ref{sec:resistivity} we focus on the intermediate regime and give the arguments for the quasi-linear resistivity in $D=3$. 
  In section \ref{sec:summary} we give a summary of the thermodynamics and transport in this model.
  Finally, in section \ref{sec:conclusions},
   we present the conclusions and  a criticism of  the work.

   The Appendices are devoted to some technical details. Precisely, in Appendix \ref{app:MF} we give the calculation of the integrals used at the mean-field approximation. In appendix \ref{app:vertices}, we give the details of the evaluation of the vertices,
    which justify the Eliashberg treatment. In appendix \ref{app:IL} we give an alternative, very simple derivation of the IL composition rules. In appendix \ref{app:polarization} we give the evaluation of the bosonic polarization, used in the form of the gauge propagator as well as in the holon  transport lifetime. In appendix \ref{app:WI}, we give a field theoretic treatment of  the Ward Identities  of this gauge theory,  relating any p-legs vertex to a (p-1)-legs one. We show as well in the most general way how the gauge symmetry protects the masses of the gauge fields in the Coulomb phase and how the mass is generated in the Higgs phase.
In appendix \ref{app:masses} we give a direct check of the cancellation of the mass in the Coulomb phase, at the first order in the perturbation theory.   Last, in Appendix \ref{app:imsigma} we give the derivation of  ${\rm Im} \Sigma_c$ and particularly show that the logarithm in $D=3$ has a thermal origin.

\section{ The model } 
\label{sec:model}

 In order to study the Mott localization of the f-electrons in the Anderson lattice, we must
 first allow them to disperse. A small dispersion of the f-electrons is naturally  
present in the most physical models for  heavy fermions.  The reason why it is  
rarely included in the starting hamiltonians\cite{millis-lee}  is that,  in the
 study of  the heavy Fermi liquid phase, the f-electrons  dispersion is irrelevant   
compared to the formation of the Kondo singlets, and one can safely approximate the narrow  
band by a flat one. 
We thus start with the Anderson lattice model, with a small dispersion
of the f-band \bea \label{eqn1} H & = & \sum_{ i, j 
\sigma} \left ( c^\dagger_{i \sigma} t_{ij} c_{j \sigma} + {\tilde
f}^\dagger_{i \sigma} ( \alpha t_{ij} + E_0 \delta_{ij}) {\tilde
f}_{j \sigma} \right )
 \\
& + & \sum_{i, \sigma} \left ((V {\tilde f}^\dagger_{i \sigma}
c_{i \sigma} + h.c.)   + U {\tilde n}_{f, i}^2 + U_{fc} {\tilde
n}_{f, i} n_{c,i} \right ) \nonumber \ , \eea where $\alpha $ is a small parameter, $\sigma$ is
a spin index belonging to the SU(2) representation, 
$t_{i j}=t$ is the hopping term taken as a
constant, $V$ is the hybridization between the f- and c- bands,
$E_0$ is the energy level of the f-electrons. ${\tilde
n}_{f,i}=\sum_\sigma {\tilde f}^\dagger_{i \sigma} {\tilde f}_{i
\sigma}$ and $n_{c,i}=\sum_\sigma  c^\dagger_{i \sigma} c_{i \sigma}$ are the operators describing the particle
number. We first study (\ref{eqn1}) in the limit of very large  on-site 
Coulomb repulsion U.  U$_{fc}$ accounts for the Coulomb interaction between the
f and c electrons; although 
$U_{fc} \ll U $  it must be taken into account in the derivation of the effective 
 theory. Here we go beyond the treatment of \cite{cath} and include  as well in the model
the RKKY interactions mediated by the conduction electrons.
The starting hamiltonian  writes
\bea \label{eqn2} H & = & \sum_{ i, j 
\sigma} \left ( c^\dagger_{i \sigma} t_{ij} c_{j \sigma} + {\tilde
f}^\dagger_{i \sigma} ( \alpha t_{ij} + E_0 \delta_{ij}) {\tilde
f}_{j \sigma} \right )
 \\
 & + & \sum_{i, \sigma} \left ((V {\tilde f}^\dagger_{i \sigma}
c_{i \sigma} + h.c.)  \right ) \nonumber \\
 & + &  \sum_{ \langle  i j \rangle } \left (  J_0  (  {\bf \tilde S}_{ f , i }
 \cdot {\bf \tilde S}_{{ f}, j} - \frac{n_{f i} \  n_{f  j} }{4} ) \right )  \nonumber \\
    & + &  \sum_{ \langle  i j \rangle } \left (  J_{RKKY} (  {\bf \tilde S}_{ f , i }
 \cdot {\bf \tilde S}_{{ f}, j}  +  J_1   {\bf \tilde S}_{ f, i } \cdot {\bf S}_{c, j}  \right ) \nonumber \ ,  \eea
where  $J_0$, $J_1$ are determined by second order perturbation theory in large 
$U/(\alpha t)$ and $U_{fc}/(\alpha t)$  and $J_{RKKY}$ is determined by second order
 perturbation theory in small $V/ D$  where D is the bandwidth of the conduction electrons.
One obtains $J_0 = 2 ( \alpha t ) ^2 / U$, $J_1= 2 ( \alpha t ) ^2 / U_{fc}$ 
and $J_{RKKY} = \rho_0 V^2 $ with $\rho_0 $ the density of states of the conduction 
electrons.  $ {\bf \tilde S}_{f }= \sum_{\alpha  \beta} { \tilde f}^\dagger_{\alpha} 
{\vec \sigma}_{\alpha \beta} {\tilde  f}_\beta $ (resp. $ {\bf S}_{c }  = 
\sum_{\alpha  \beta} c_{\alpha}^\dagger {\vec \sigma}_{\alpha \beta} c_\beta$) and ${\vec \sigma}$ 
is the Pauli matrix.    The summation over $\langle i j \rangle$  can  run on first 
nearest neighbors as well as second, third etc... nearest neighbors.  Here we simplify
 the discussion by restricting  all spin interactions to first nearest neighbors 
only, hence restricting our attention to the commensurate AF case. It has to be kept 
in mind that this is not the most  generic situation and that frustration naturally appear  in this model.

  We then take the  $U \rightarrow \infty$ limit of (\ref{eqn2}). We
account for the constraint of no double occupancy through a
Coleman\cite{coleman84}  set of bosons $ (b^\dagger, b) $
enslaved to a constraint on each site $ \sum_{\sigma} f^\dagger_{i
\sigma} f_{i \sigma} + b^\dagger_i b_i = 1 $.  We make  at each site the transformation 
\beq \label{eqn40} {\tilde f}_{i \sigma} \rightarrow  f_{i \sigma} b^\dagger_i \  , \eeq 
where  the $f^\dagger$- creation operators are called  ``spinons'' and the $b^\dagger$ ``holons''.   The physical  electron ${\tilde f}$ 
splits into a ``holon'' and a ``spinon'' under the  transformation (\ref{eqn40}).
 We notice that, upon this
transformation, the slave boson drops out of  all bi-linear products of fields at the same site.   Indeed, one can  explicitly show  that at site ``i'':
\beq \left \{ \begin{array}{l}
f^\dagger_{i \sigma} b_i b^\dagger_i  f_{i \sigma} |0 \rangle  =  f^\dagger_{i \sigma} f_{i \sigma}  |0\rangle  = 0 \\
f^\dagger_{i \sigma} b_i b^\dagger_i  f_{i \sigma} |\sigma \rangle =  f^\dagger_{i \sigma} f_{i \sigma} |\sigma \rangle  = n_{f, i \sigma}  |\sigma \rangle \end{array} \right  .  \eeq Another way to see this is to apply the constraint inside the operator.
\bea
f^\dagger_{i \sigma} b_i b^\dagger_i  f_{i \sigma}  & = &  ( 1- n_b) f^\dagger_{i \sigma}  f_{i \sigma}  \nonumber \\
& = & n_{f i} \   f^\dagger_{i \sigma}  f_{i \sigma} \nonumber \\
& = &  f^\dagger_{i \sigma}  f_{i \sigma} \ . \eea The last line is obtained by observing that the constraint of no double occupation is equivalent to  $n_{f i }^2= n_{f i }$.

The effective Lagrangian is then \bea \label{eqn3} {\cal L}
& = & \sum_{ i, j \sigma} \left ( c^\dagger_{i
\sigma} ( \partial_\tau \ \delta_{ij} + t_{ij} )  c_{j \sigma} \right .  \\
& + & \left . f^\dagger_{i \sigma}
(  b_i \alpha t_{ij} b_j^\dagger + (\partial_\tau + E_0 +
\lambda_i ) \delta_{ij} ) f_{j
\sigma} \right )\nonumber \\
& + &   \sum_i b^\dagger_i \left ( \partial_\tau + \lambda_i \right
) b_i
-\lambda  + \sum_{i, \sigma}  ( V f^\dagger_{i \sigma } b_i c_{i
\sigma } + h.c. ) \nonumber   \\
& + &    \sum_{ \langle  i j \rangle } \left ( J   {\bf  S}_{ f , i } \cdot {\bf  S}_{ f, j}  +  J_1   {\bf  S}_{ f, i } \cdot {\bf S}_{c, j}  \right ) \nonumber \ , \eea where  the
 constraint has been implemented through a Lagrange multiplier
 $\lambda_i$.
 Note that the  spin operator $ {\bf  S}_{ f }= \sum_{\alpha  \beta} { f}^\dagger_{\alpha} {\vec \sigma}_{\alpha \beta} {  f}_\beta $ is now expressed solely in terms of the spinons and thus is insensitive to the slave bosons.
 
In  the following of the paper we consider a ``large N'' extension of the Lagrangian (\ref{eqn3}) 
by enlarging the spin group from SU(2) to SU(N). The indices $\sigma$ now belong to the SU(N) group.

 \section{Mean-field approximation } 
  \label{sec:MF}
 
\begin{figure}[h]
\includegraphics[width=2.6in]{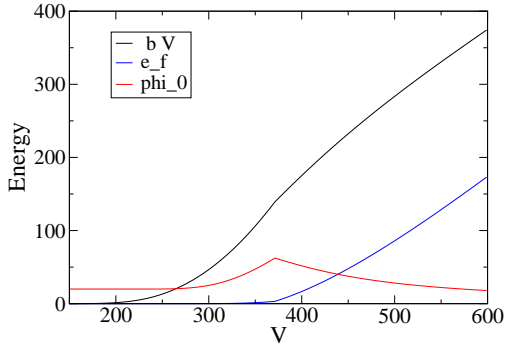}
\caption{ Effective hybridization $V b$ (black curve), the f-band chemical potential
$\epsilon_f= E_0 + \lambda$ (blue curve) and  $20 \phi_0$ (red curve), as a function of V. The electron
bandwidth is $D= 1000$. The chemical potential $\mu = 0$, the
ratio of f- and c- masses is $\alpha =0.1$. $\phi_0$ is evaluated self-consistently around finite RKKY value of
$10^{-3} D$.
The f-energy level is $E_0 = - 500$. The mean field equations are solved for $N=2$.  The effect of the field $\sigma$ is negligible, 
hence not shown in the figure. } \label{fig1}
\end{figure}

  In the mean-field approximation, we minimize the effective action with four fields : ($ \lambda $, $b $,   $ \phi_0$,  $\sigma_0$ ), where
  a static  and uniform approximation is made on all fields. $\phi_0 $ is the  uniform spin liquid parameter, which decouples  the short range AF interaction
  $  J   {\bf  S}_{ f , i } \cdot {\bf  S}_{ f, j}  \rightarrow  \phi_0 \sum_{i, \sigma} ( f^\dagger_{i \sigma} f_{i \sigma} + h.c.)  - N \phi_0^2/ J $.
  $\sigma_0$ is the uniform field which decouples the induced Kondo-like interaction
  $J_1   {\bf  S}_{ f, i } \cdot {\bf S}_{c, j} \rightarrow \sigma_0 \sum_{ i , \sigma}  ( f^\dagger_{i \sigma} c_{i \sigma} + h.c.  ) - N\sigma_0^2/J_1 $. 
    The minimization of the free energy leads to the following mean-field equations:
    \bea \label{eqn5}
    & & T \sum_{k, \sigma, \omega_n} \alpha \epsilon_k G_{ff}+ \frac{V}{b} \sum_{k, \sigma, \omega_n} G_{fc} + \epsilon_f - E_0 = 0 \ ;\nonumber \\
    & & T \sum_{k, \sigma, \omega_n} G_{fc} + N \sigma_0/ J_1 = 0  \ ; \nonumber \\
    & & T \sum_{k, \sigma, \omega_n }    ( \epsilon_k / D )  G_{ff}  + N \phi_0 / J = 0 \ ; \nonumber \\
    & &  b^2 + T \sum_{k, \sigma, \omega_n}  G_{ff} = N/2  \ , \nonumber \eea
where  $\epsilon_k$ is a typical dispersion of the conduction electrons, $\epsilon_f= E_0 + \lambda$ and the dispersion of the spinon band  is taken to be $\epsilon_k^0 =\left ( \alpha b^2 + (\phi_0/ D) \right ) \epsilon_k + \epsilon_f $.
We have the following Green's functions
\beq \begin{array}{l}
\displaystyle {G_{ff} = \frac{i \omega_n - \epsilon_k }{ ( i \omega_n -\epsilon_k ) ( i \omega_n - \epsilon_k^0 ) - (V b + \sigma_0)^2} } \  , \\
\\
\displaystyle{ G_{fc} = \frac{ V b + \sigma _0 }{ ( i \omega_n -\epsilon_k ) ( i \omega_n - \epsilon_k^0 ) - (V b + \sigma_0)^2} } \ , \\
\\
 \displaystyle {G_{cc} = \frac{ i \omega_n - \epsilon_k^0 }{ ( i \omega_n -\epsilon_k ) ( i \omega_n - \epsilon_k^0 ) - (V b + \sigma_0)^2} } \end{array} \ . \eeq The mean field equations are solved in the case of a linearized bandwidth and for $N=2$; energy scales extracted from the mean -field studies are as well written for N=2. 
 The summations over $(k, \omega_n)$ can then be performed analytically and are given in Appendix \ref{app:MF}; the set of  resulting equations is then solved numerically.  Since the interaction $ J_1   {\bf  S}_{ f, i } \cdot {\bf S}_{c, j}$ generates some additional  Kondo coupling, it is not obvious  that a Mott transition- a point in the phase diagram where $b=0$ - occurs at finite V.  However this is what happens, the additional Kondo coupling $ \sigma_0$ being itself driven to zero at the transition.
 The result is displayed in Fig.\ref{fig1}.
 
     We can  study analytically the nature of the the fixed point by expanding the mean-field equations (\ref{eqn5}) around 
     \beq \begin{array}{l}
     \sigma_0 =0 \ ; \\
     b=0 \ ; \\
     |\phi_0 | = c J; \;    \mbox{with} \;  c \simeq 0.1 \ . \end{array}
 \eeq From  the first equation, we get that 
 \beq \begin{array}{ll}
 &  \Pi_{fc} (0)  +  (\epsilon_f - E_0)/ V^2 \simeq 0 \;  \\
 \\ \mbox{where} \; \;  & {\displaystyle \Pi_{fc} (0) =   T \sum_{k, \omega_n, \sigma} \frac{1}{( i \omega_n - \epsilon_k ) ( i \omega_n - \epsilon_k^0 ) } } \ .  \end{array} \eeq   Using, for the conduction electrons and the spinons,  linearized dispersions of the form
  $ \epsilon_k = v_F ( k - k_F) $ , $\epsilon^0_k = v_0  ( k - k_0) $   with $ v_0 = 2 \phi_0 / k_F $ one obtains
 \beq
 \Pi_{fc} (0) =   N \rho_0 \frac{{\rm Log} ( \alpha^\prime ) }{ ( 1 - \alpha^\prime ) } \  , \eeq
  \beq \mbox{with} \hspace{0.5 cm}  \alpha^\prime = \alpha b^2 + \phi_0 / D \ .  \eeq One can convince oneself that the  fixed point occurs when
 $  \phi_0 =  D Exp \left [  \frac{ E_0}{  N \rho_0 V^2 } \right ] $. One recognizes here the typical Kondo scale of the problem\cite{andersonKondo,nozieres}.   Hence 
 \beq
 J_{crit} = D {\rm Exp } \left [  \frac{ E_0}{ N \rho _0 V^2 } \right ]  \ .\eeq  One sees that the  location of the QCP in the phase diagram depends crucially on the spin liquid parameter $J$.  Moreover,  it is important  for the stability of the  mean -field equations that the bandwidth stays finite when $ b \rightarrow 0 $. This is achieved   with the uniform spin liquid parameter  $\phi_0 $ being finite  at the transition.
 The mean field equations are of little help to determine the nature of the localized phase where $b=0$.  In the simplified scheme we have taken, this phase is a uniform spin liquid - $\phi_0$ being finite and uniform.   The spinons can however have some other symmetries, and  it would be useful, for example, to determine the location of the transition when spinons with a nodal Fermi  surface are used. 
  One can as well allow for  AF order in the  localized side and study the stability of the phase diagram. This  is the program for future work.
  
  \section{Modulations of the order parameter}
  \label{sec:modulations}
  
   In the previous section,we have naturally considered that the order parameter $b$ was ordering at $q=0$. This issue has to be re-considered  keeping in mind the  nature of the spinon  Fermi surface.  In  this section we review for completeness, the results obtained in \cite{us}.
     At the QCP, the effective mass of the order parameter writes
     \bea
     D_b^{-1} (q,0)  &  =  &  m_b \  ; \nonumber \\
     m_b  (q)  &  =   &  \rho_0 ( - E_0  +  V^2  \Pi_{fc } (q) )  \ , \eea  where $\Pi_{fc} (q) $  is the static $fc$-polarization, taken at  finite momentum $q$ but zero frequency. At  the QCP , the minimum of the effective mass determines the ordering wave vector.  Two situations are  to be considered. First, if both the f- and c- bands are electron-like,   $\Pi_{fc} (q ) $  obtains its minimum at $q_0=0$. Note that in that case,  the curvature of the Fermi surface has to be included to see the minimum.
     Second, if  one band is electron-like but the other one  is hole-like, then the situation is analogous to  the FFLO ordering in  superconductors \cite{FFLO} or to change density waves in Chromium \cite{rice}. At the QCP, the ordering wave vector is at
     $q_0\simeq 1.2  q^* $ where
     \beq \label{eqn6}
     q^* =  | k_F - k_0 |    \  , \eeq is the difference between the Fermi wave vectors of the two species.
     The determination of the ordering wave vector inside the ordered phase is much more involved 
than at the QCP. It led to  a full literature in the case of  FFLO superconductivity\cite{FFLO}. One generally expects 
a first order transition towards a uniform order inside the ordered phase.   Here the situation is 
rather more complex than in the FFLO case because the order parameter $b$ carries one quantum of 
  gauge charge.  Gauging out the theory, even at the mean field level, is required to deduce  the observable 
quantities. This  study definitely deserves more  work.  Particularly,  it would be  interesting to 
see what kind of superconductivity occurs in a Kondo phase where the hybridization is modulated in space.

    \section{ Amplitude fluctuations}
    \label{sec:amplitude-fluc}
\begin{figure}[h]
\includegraphics[width=2.6in]{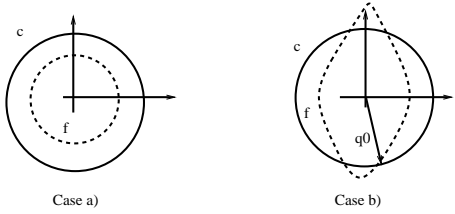}
\caption{ Illustration of the two typical cases of interest. In case a) the two Fermi surfaces  of the spinons and conduction electrons  are centered and there is a gap between them; in case b) the two Fermi surfaces intersect.  } \label{Fermisurfaces}
\end{figure}

\begin{figure}[h]
\includegraphics[width=2 in]{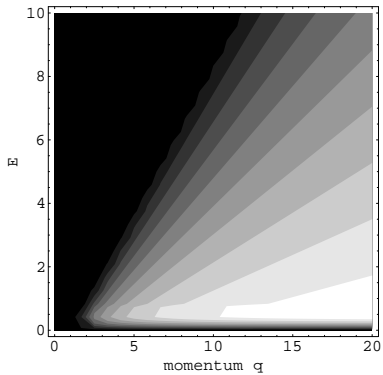}
\caption{ Spectral weight of the $\Pi_{fc}$ polarization in the $(q, \omega)$-plane. We have taken $q^* = 1$, $v_F=1$, $\alpha^\prime = 0.01$.   We see that the weight is concentrated  in the regime II ($z=3$) for $q \geq q^*$, where  one has particle-hole damping.} \label{spectralweight}
\end{figure}

     In the two following sections, we describe the random phase approximation (RPA) evaluation of the amplitude and gauge fluctuations.    The fluctuations are studied in the case of 
 the order parameter  condensing at $q=0$. 
     The fluctuations of the amplitude of  the order parameter are more complex 
than what was considered in \cite{ssv} since the order parameter is coupled to two types of fermions: 
the f-spinons and the conduction 
electrons. Neglecting for the moment the effect of gauge fluctuations,  within the RPA the polarization is similar to a Lindhard function, 
but with two different types of fermions.
\vspace{- 30 pt}
\begin{center}
$ \Pi_{fc} = $ \begin{picture}(90,70)(0,90)
\Text(8,99)[c]{$b$}
\Text(35,113)[c]{$c$}
\Text(35,63)[c]{$f$}
\LongArrow(6,85)(12,85)
\Vertex(20,90){1.5}
\Vertex(50,90){1.5}
\Photon(5,90)(20,90){2}{2.5}
\Photon(50,90)(65,90){2}{2.5}
\DashArrowArcn(35,90)(15,0,180){1.5}
\ArrowArcn(35,90)(15,180,0)
\end{picture}
\end{center}
\vspace{ 30 pt }
 \beq \label{eqn11}
D^{-1}_b (q, i \Omega_n) = \rho_0  \left [ -i \Omega_n + \  \delta  + a q^2 +  \Pi_{fc} ( q, i \Omega_n) \right ] \ , \eeq with
$ a= - N {\rm Log } ( \alpha^\prime )/ [ ( 1 - \alpha^\prime )^2 k_F^2 ] $, the bosonic mass $ \delta = - E_0 $ comes from the mean-field equations,  and $\rho_0$ is the density of states of the conduction electrons. Note that, in our definition, $\Pi_{fc}$  is the ``dynamical'' polarization, corresponding to the  infra-red (IR) sector. In the text, we use this definition for every polarization.  Let's here evaluate $\Pi_{fc}$.
There are two cases of interest, as shown in Fig.\ref{Fermisurfaces}.    Depending whether the Fermi surface of the spinons and conduction electrons intersect or not, we have two forms for the amplitude propagator.
Let's start with case a) where  there is a gap between the spinon and electron  Fermi surfaces.
Using linearized bands, the polarization can be computed analytically and we get at $T=0$:
\bea \label{eqn12}
 & & \Pi_{fc} ( q, i \Omega_n ) = \frac{N \rho_0}{4 \pi } \int  \ d \epsilon \ d  \omega \ d cos \theta 
\times \nonumber \\
& & \frac{1}{ ( i \omega + i \Omega_n  - \epsilon  - v_F q \ cos\theta )  
( i \omega - \alpha \epsilon - \alpha v_F q^* ) } ; \nonumber \\
 & = & \frac{ - \rho_0} { 2 v_F \alpha^\prime q ( 1 - \alpha^\prime )  }   \times \\
 & & \left [  ( - \alpha^\prime i \Omega_n + \alpha^\prime v_F ( q-q^*) ) 
Log( - i \alpha^\prime \Omega_n  + \alpha^\prime v_F ( q-q^*) )  \right . \nonumber \\
 & - & ( - \alpha^\prime i \Omega_n + \alpha^\prime v_F ( -q-q^*) ) 
Log( - i \alpha^\prime \Omega_n  + \alpha^\prime v_F ( -q-q^*) ) \nonumber \\
& - &  ( -  i \Omega_n + \alpha^\prime v_F ( q-q^*) ) 
Log( - i  \Omega_n  + \alpha^\prime v_F ( q-q^*) ) \nonumber \\
& + & \left . ( -  i \Omega_n + \alpha^\prime v_F ( -q-q^*) ) 
Log( - i  \Omega_n  + \alpha^\prime v_F ( -q-q^*) )  \right ] \nonumber \ . \eea
We see that  an additional scale $q^*$ naturally emerges from the polarization. $q^*$ is the difference between the
  Fermi wave vectors of the f-spinons and c-electrons.  
Here we have taken both the  spinon Fermi surface and the electron Fermi surface
to be spherical and centered  with respect to each other, hence $q^*$  isotropic. We define the energy scale for $N=2$ by 
\beq \label {eqn13}
E^* = 0.1 \alpha^\prime  D \left ( \frac{ q^*}{k_F} \right ) ^3 \ . \eeq 
We can expand (\ref{eqn12}) in four different regimes.
(i)  For  $ q \leq q^*, \; |\Omega_n | \leq E^* $  we have \beq \label{eqn14}D^{-1}_b (q, i \Omega_n) \simeq   \rho_0  \left 
[  {\tilde \delta } + a q^2 - N \frac{ i \Omega_n}{\alpha^\prime v_F q^* }  \right ],  \  \eeq
 with $ {\tilde \delta } = - E_0 + \rho_0 V^2 Log \alpha^\prime / ( 1 - \alpha ^\prime )  $.  
 (ii) For  
 $ q \leq q^*, \;  E^* \leq  |\Omega_n | $  we have
\beq D^{-1}_b (q, i \Omega_n) \simeq   \rho_0  \left 
[  {\tilde \delta } + a q^2 +N \frac{ {\rm Log }  | \Omega_n| }{\alpha^\prime v_F q^* }  \right ]  \ . \eeq
(iii) For $ q \geq q^*, \; \alpha ^\prime v_F q \leq |\Omega_n| \leq  v_F q$  we have 
\beq D^{-1}_b (q, i \Omega_n) \simeq  \rho_0 \left 
[  {\tilde \delta } + a q^2 + N \frac{ {\rm Log } | \Omega_n| }{\alpha^\prime v_F q }  \right ]  \ . \eeq
(iv ) For $ q \geq q^*, \;  v_F q \leq |\Omega_n| $  we have 
\beq \label{eqn15}  D^{-1}_b (q, i \Omega_n) \simeq  \rho_0 \left 
[  {\tilde \delta } + a q^2 + N \frac{  | \Omega_n| }{\alpha^\prime v_F q }  \right ]  \ . \eeq
Note the peculiar form of the boson propagator, where  the different sectors are not in the same footing in large N\cite{note9}.  We don't know any way to avoid this problem within the  Eliashberg theory, and that's why we call it a large N {\it expansion} rather than  a large N {\it limit}. We refer the reader to \cite{aim,jerome} for further details about this and to next section, just mentioning here that the merit of the Eliashberg theory is to  capture all the frequency divergences  within the leading and second leading orders in $1/N $.
 To each regime one can associate a different dynamical exponent. In regime (i) the dynamical exponent is $z=2$ and  the propagator corresponds 
to an un-damped bosonic mode. In regime (ii) the dynamical is exponent $z= \infty $. In regime (iii) we get $z=1$. 
Finally, regime (iv) has $z=3$. It was shown that the spectral weight  in the $(\Omega, q)$-space is most entirely centered 
in the regime (iv) where $z=3$\cite{us}. This feature is illustrated in Fig \ref{spectralweight}.
  Physical arguments enable to simplify the discussion.   At low momentum and low energy, the
particle-hole continum is gapped, due to the mismatch of the two Fermi surfaces.  There is propagation of a single boson mode
with exponent  $z=2$ (\ref{eqn14}). At high momentum and high energy, one can neglect the mistmatch of the two 
Fermi surfaces; the two fermion species  behave as if they were identical.  The polarization behaves like a 
Lindhard function (\ref{eqn15}) 
with $z=3$; this form is typical of a $q=0$ phase transition.

 As we have shown above, an energy scale $E^*$ is present in the  quantum critial regime, separating two regimes with 
universal exponents.  In the low energy regimes $T \ll E^*$  the  fluctuations of the amplitude are gapped and 
 thermodynamics is dominated  by  gauge fluctuations.   For $T\gg E^*$ 
 the thermodynamics is  in the universality class of the ferromagnet with $z=3$.

  On can ask the question:   -``why is $E^*$ so low ?''
To answer it we give an estimate of  $E^*$. One can first identify  $\alpha^\prime D$ with the temperature  $T_0$ at which
the entropy $R {\rm  Ln} 2$ is released in the Anderson lattice. This scale is seen experimentally in every compound, as 
a bump into the thermodynamic and transport observables.   For $ T \geq \alpha^\prime D $, the  spinons loose their dispersion; hence  one can  consider that they 
behave as free spins and that the entropy $R Ln 2 $ is released. This  observation fixes the scale $T_0$ to be roughly $ 24 K$ for a compound like
YbRh$_2$Si$_2$ and $ 50 K $ for CeCoIn$_5$. We see from Eqn.(\ref{eqn13}) that even for very large $q^*$, $E^*$ is already an
 order of magnitude lower than $T_0$, which makes it a  small scale already. Next, depending on the form of the spinon Fermi surface, 
$q^*$ can be as small as $ 0.1 k_F$, in which case the scale $E^*$  is of the order of   $ 10 mK$; 
which means it is  not experimentally reachable. The fact that $E^*$ relies on the imponderable form of  the spinon Fermi surface is assuredly the weak point
of  this theory.

\begin{figure}[h]
\includegraphics[width=1.0in]{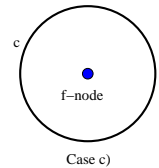}
\caption{ The third case of interest, where the spinon f Fermi surface is nodal, like for  the  U(1) gauge theory of the one band Hubbard model, in $D=2$. This is in the same universality class as the case a)  } \label{Fermisurfaces2}
\end{figure}

 To illustrate  this point, we look at  the  more generic case b), where the spinon Fermi surface and the electron Fermi 
surface intersect in 
two ``hot lines''. Since the anisotropy is broken,  we define a new $q^*$ as 
\beq q^* ={\rm  Max}_{(\theta,\phi ) }[  | k_F- k_0 |] \ , \eeq where the maximum is taken over anglular variables on the Fermi surface. 
In this case, the regime (iv) is unchanged, but the regime (i) is different, since the particle-hole continum  
is  not gapped.  The boson propagator in regime (i)   is now
\beq \label{eqn16}D^{-1}_b (q, i \Omega_n) \simeq    \rho_0 \left 
[  {\tilde \delta } + a q^2 + N c \frac{  | \Omega_n |}{\alpha^\prime v_F q_0 }  \right ],  \  \eeq where $q_0$ 
 is the modulus of the
wave  vector at which  the two Fermi surfaces intersect, and  c is a coefficent depending on the shape of the Fermi surfaces.
 We see that the propagator still has the dynamical exponent $z=2$ but it  is  now damped; hence in the same universality class as
a 3 D anti-ferromagnetic SDW  QCP. We note that the change of the spinon Fermi surface  between case a) and case b) didn't 
affect  the regime for $T \geq E^*$.  Although we cannot prove that this result is  general, we  illustrate it with considering a third case where the f-spinon Fermi surface is nodal in 2 D, as shown in Fig.\ref{Fermisurfaces2}. This case is interesting because it is the exact analogue to the description of the one band Hubbard model within the U(1) or SU(2) slave-boson gauge theory\cite{lee-review,leenagaosa}. One can show that  the amplitude propagator is formally equivalent to case a), where the two Fermi surfaces are gapped.    Hence in that case as well, the regime with $T \geq E^*$ is unaffected by the form of the spinon Fermi surface, although  $E^*$ is supposedly bigger that in case a), since  $q^*  $ and $k_F$ are of the same order of magnitude.  In conclusion, the regime $ T \geq E^*$ depends on the  existence of the spinon Fermi surface but not on its shape.

 \section{Gauge fluctuations}
  \label{sec:gauge-fluc}
  
   The order parameter of our phase transition is not  a macroscopic order parameter, but relies on the condensation of a gauge boson  carrying  one quantum of charge  of the gauge invariant theory.  We study here the U(1) local invariance of the Lagrangian. 
   For clarity we start with  the evaluation of the polarization of the gauge fields within RPA. The WI  constraining the masses are given in  Appendix \ref{app:WI}.  We start with the version of (\ref{eqn2}) after having performed the Hubbard-Stratonovich transformation.
   \bea \label{eqn7} {\cal L}
& = & \sum_{ i, j \sigma} \left ( c^\dagger_{i
\sigma} ( \partial_\tau \ \delta_{ij} + t_{ij} )  c_{j \sigma} \right . \nonumber \\
& + & \left . f^\dagger_{i \sigma}
(  b_i \alpha t_{ij} b_j^\dagger + (\partial_\tau + E_0 +
i \lambda_i)\delta_{ij} ) f_{j
\sigma} \right )\nonumber \\
& + &   \sum_i b^\dagger_i \left ( \partial_\tau + i \lambda_i \right
) b_i
- i \lambda_i \nonumber \\
& + &  \sum_{i, \sigma}  \left ( (V  b_i + \sigma_i) f^\dagger_{i \sigma }  c_{i
\sigma } + h.c. ) \right )  \nonumber   \\
& + & \frac{N}{J} \sum_{\langle i j \rangle }  | \phi_{ij} |^2 +   \frac{N}{J_1} |\sigma_i |^2 \nonumber \\
& + & \sum _{\langle i  j \rangle \sigma } \left ( f^\dagger_{i \sigma} f _{i \sigma}  \phi_{ij} + \phi _{ij} ^*  f^\dagger _{j \sigma }f_{j \sigma} \right )  \ . \eea 
   Under the gauge transformation
  \beq \label{eqn8b}
   \begin{array}{l}
   f_i \rightarrow f_i \  e^{- i \theta_i} , \\
   b_i \rightarrow b_i \   e^{i \theta_i}, \\
   \sigma_i \rightarrow \sigma_i \ e^{i \theta _i } , \\
   \lambda_i \rightarrow \lambda_i + \partial_\tau \theta_i , \\
   \phi_{ij} \rightarrow \phi_{ij} \ e^{i \theta_i - \theta_j} , \end{array} \eeq
    the Lagrangian (\ref{eqn7}) acquires  a total derivative  $ {\cal L } (\tau) \rightarrow {\cal L } (\tau ) 
 - i \sum_i  (\partial _\tau \theta_i)  $ which  is a multiple of $2 i \pi $ and thus leaves the 
 Lagrangian invariant.   The above Lagrangian is  thus invariant under a U(1) local -or gauge - symmetry.
   In the  mean-field treatment above, we have considered that  the Hubbard-Stratonovich fields 
can be taken at their saddle points
 $ \langle \sigma_i \rangle = \sigma_0 $, $ \langle \lambda_i \rangle = i \lambda $ , 
 $ \langle  b _i \rangle  = b$ and $\langle  \phi_{ij} \rangle = \phi_0  $ and we have 
neglected the fluctuations of the phases of the field as well as of the amplitudes.
  It is safe to assume that the amplitude fluctuations of  $\phi_{ij} $ are gapped since 
$\phi_0 $ doesn't  vanish  through the phase diagram. At the QCP, however, the amplitude 
fluctuations of  $b_i $ and $\sigma_i$ becomes massless.
  Another issue is the phase of the gauge fields. In order to study them, it is convenient to 
 take the continuum limit in  the Lagrangian (\ref{eqn7}).   For this, we write 
$ \phi_{ij} = \phi_0  e ^{i  {\bf a}_{ij} } $ and now the second and third  terms in Eqn.(\ref{eqn7}) describe 
spinons and holons in the presence of the  fictitious electromagnetic field  ${\bf B}$  associated with the 
 vector potential  ${\bf a}$ with $ \int_{r_i}^{r_j} {\bf a} \cdot d {\bf l} = {\bf a}_{ij} $
 \cite{lee-review,leenagaosa}.
  We then coarse grain the Lagrangian (\ref{eqn7}) to obtain the continuous limit
\bea \label{eqn8} 
  {\cal L}( {\bf r}, \tau ) & = & \sum_\sigma \int d {\bf r }  c^\dagger_{\sigma}   \left ( \partial \tau  - \frac{\nabla ^2}{ 2 m } - \mu \right ) 
   c_\sigma \nonumber \\
   &  + & f^\dagger_\sigma   \left ( \partial \tau - \frac{ (\nabla  + i e \ {\bf a }/ c  ) ^2}{ 2 m_f}  + \epsilon _f  + i a_0  \right ) f_{\sigma} \nonumber \\
   & + &  b^\dagger   \left ( \partial \tau - \frac{ (\nabla  + i  e \ {\bf a} / c  ) ^2}{ 2 m_b}   + \lambda + i a_0   \right ) b  \nonumber \\
   & + & \frac{N}{J_1} \sigma({\bf r } )^2 + \frac{N}{J } \phi_0^2  \nonumber \\
   & + &  \sum _\sigma \int d {\bf r } \left [ ( V b + \sigma )   f^\dagger_\sigma   c_\sigma + h. c. \right ]  \ . \eea
    When the  order parameter  $  b + \sigma/V $ condenses, the gauge field ${\bf  a} $ acquire a gap proportional to 
      $  (b + \sigma/V)$ according to the rules  for  the condensation of a Higgs boson\cite{peshkin}. This phenomenon is  responsible for the Meissner effect in the case of a superconductor. In Appendix \ref{app:WI} we derive the generation of the mass using the 
       Ward identities  associated with the gauge invariance.
       
        A non trivial issue is the gauging out of the theory.   The  slave-boson representation required the use of additional fields, hence the field theory is redundant and some gauge fluctuations can be factorized out of the partition function.   Since one deals with a U(1) gauge  theory, one gauge fixing constraint only is required. When gauging out the theory, it is convenient to chose the radial gauge  for the Higgs  phase - also called  here  ``heavy Fermi liquid'', which follows previous studies  of the Kondo impurity \cite{read,coleman84} and of the Kondo lattice\cite{levin}. Namely, we choose the order parameter to be real, with for example  \beq  b + \sigma / V = | b + \sigma / V |  \  . \eeq This constraint is  alternatively called  the ``physical gauge'' in field theory, since the gapping of the gauge fields in the Higgs phase is transparent in this gauge \cite{peshkin}.  
  When  the order parameter  vanishes, in the so called Coulomb phase,  it is clever to use the  Coulomb gauge  imposing the condition 
  \beq  \label{eqn43}
  \nabla \cdot {\bf a } = 0 \ , \eeq so that the vector fields become purely transverse.  In the Coulomb gauge, the  scalar fields ($\mu=0$) and vector fields decouple; the scalar fields are massive, which are nothing  than the density-density correlation function,   while the mass of the vector fields remain massless(see Appendix \ref{app:WI}).  The polarization of the vector fields gives rise to a Reizer singularity \cite{reizer} described below. Of course, formally  all gauge fixings should be equivalent, and there is no  deep reason to chose one constraint rather than the other.  However,  if  for example one tries to  work  with the radial gauge in  the  Coulomb phase, lots of divergences appear which  have to cancel  at the end, for physical quantities. The reason is that radial coordinates are ill-defined when  the radius vanishes (here  the  radius is the modulus of the order parameter $  | b  + \sigma /V |$ ) . Some divergences are then hidden in the Jacobian  of the transformation.   We are not aware of any  field theoretic treatment of this case for condensed matter systems.   
  
  At the QCP we thus work in the Coulomb gauge  where the  vector fields have the purely transverse form :
  \beq
   D_{ij}^{-1} ( q, i \Omega_n )  =     \Pi( q, i \Omega_n )  ( \delta_{ij}  - q_i q_j / q^2 ) \  ,  \mbox{with} \eeq
   \bea \label{eqn9}   \Pi(q, i  \Omega_n )   & = &    - \frac{1}{2}  \langle  T_\tau [  {\bf J}_{f,i} ( {\bf r }, \tau )  {\bf J}_{f,j} ( 0,0) ] \rangle   \nonumber \\
   & + &  \delta_{ij} \frac{\rho_0}{2 m_f}  \delta ({\bf r}) \delta (\tau )  \nonumber \\
   & - &   \frac{1}{2}  \langle  T_\tau [  {\bf J}_{b,i} ( {\bf r }, \tau )  {\bf J}_{b,j} ( 0,0) ] \rangle \nonumber \\
   & + &   \delta_{ij} \frac{\rho_0}{2 m_b}  \delta ({\bf r}) \delta (\tau )  \  . \eea   The first term in (\ref{eqn9}) corresponds to the paramagnetic contribution while the second term is the diamagnetic part. ${\bf J}_f $ is the current operator defined as  $ {\bf J}_f = i/ ( 2 m_0) [ f^\dagger \nabla f - ( \nabla f^\dagger) f ] $  and  $ {\bf J}_b = i/ ( 2 m_b) [ b^\dagger \nabla b - ( \nabla b^\dagger) b ] $  After Fourier transforming, the polarization can be written as
   \bea 
 & &   \Pi_{ij} ( q,  i \Omega_n )   =  \frac{T }{ 2 m _f^2} \sum_{k,  \sigma, \omega_n } ( {\bf k} + {\bf q}/2)_i ( {\bf k } + {\bf q} /2)_j  \nonumber \\
   & \times & G_{ff} ( {\bf k } + {\bf q}, i \omega_n  + i \Omega_n )  G_{ff} ( {\bf k }, i \omega_n) + \delta_{ij} \frac{\rho_f}{2 m_f} \nonumber \\
   &+& \frac{T }{ 2 m _b^2} \sum_{k,  \omega_n } ( {\bf k} + {\bf q}/2)_i ( {\bf k } + {\bf q} /2)_j  \nonumber \\
   & \times & G_{b} ( {\bf k } + {\bf q}, i \omega_n  + i \Omega_n )  G_{b} ( {\bf k }, i \omega_n) + \delta_{ij} \frac{n_b}{2 m_b}  \nonumber  \  . \eea 
   The  vector field propagator has the form
   \beq \label{eqn10}
    D_{ij}^{-1} ( q, i \Omega_n )  =   \left (  \Pi_f + \Pi_b \right )  ( \delta_{ij}  - q_i q_j / q^2 ) \  , \eeq 
    where  at the QCP and at $T=0$,
    \bea \label{eqn10b}
    \Pi_f (q, i \Omega_n )  & = &  \frac{N }{2   m_f} \left [ \frac{ \pi | \Omega_n |}{v _F q } +  (q/k_F)^2  \right  ]  \nonumber \\
    \Pi_b ( q, i \Omega_n)  & = &  \frac{1}{2 m_b} \left [ \frac{ \pi f_d  |\Omega_n|}{ q} + (q^2)/ ( 2 m_b)  \right ] \  , \eea   with $ f_d= \int  ( q^2/ d) d^{(d-1)} q  f_0(q)/( 2 \pi )^{d-1} $and $f_0(q)$ is a UV cut-off\cite{note9}. The evaluation of $\Pi_b$ is given in Appendix \ref{app:polarization}.  Pictorially, the polarization can be written
   \vspace{- 30 pt}
   \begin{center}
   $ \Pi_f + \Pi_b $ =  \begin{picture}(90,70)(0,90)
\Text(8,100)[c]{${\bf a}_\mu$}
\Text(59,100)[c]{${\bf a}_\nu$}
\Text(35,115)[c]{$f$}
\Text(35,63)[c]{$f$}
\LongArrow(6,85)(12,85)
\LongArrow(55,85)(60,85)
\Vertex(20,90){1.5}
\Vertex(50,90){1.5}
\ZigZag(5,90)(20,90){2}{2.5}
\ZigZag(50,90)(65,90){2}{2.5}
\DashArrowArcn(35,90)(15,0,180){1.5}
\DashArrowArcn(35,90)(15,180,0){1.5}
\Text(75,90)[c]{$+$}
\Text(90,100)[c]{${\bf a}_\mu$}
\Text(140,100)[c]{${\bf a}_\nu$}
\Text(115,115)[c]{$b$}
\Text(115,63)[c]{$b$}
\LongArrow(86,85)(92,85)
\LongArrow(134,85)(140,85)
\Vertex(100,90){1.5}
\Vertex(130,90){1.5}
\ZigZag(85,90)(100,90){2}{2.5}
\ZigZag(130,90)(145,90){2}{2.5}
\PhotonArc(115,90)(15,0,180){2}{6}
\PhotonArc(115,90)(15,180,0){2}{6}
\LongArrowArc(115,90)(10,60,120)
\end{picture}
\end{center}
\begin{center}
$+$ \begin{picture}(50,50)(0,28)
\Text(8,15)[c]{${\bf a}_\mu$}
\ZigZag(5,20)(20,30){2}{2.5}
\Vertex(20,30){1.5}
\ZigZag(20,30)(35,20){2}{2.5}
\PhotonArc(20,40)(10,0,360){2}{10}
\LongArrowArc(20,40)(5,60,120)
\Text(30,15)[c]{${\bf a}_\nu$}
\end{picture} $ + $ \begin{picture}(50,50)(0,28)
\Text(8,15)[c]{${\bf a}_\mu$}
\ZigZag(5,20)(20,30){2}{2.5}
\Vertex(20,30){1.5}
\DashArrowArc(20,40)(10,0,360){1.5}
\ZigZag(20,30)(35,20){2}{2.5}
\Text(30,15)[c]{${\bf a}_\nu$}
\end{picture}
\end{center}
\vspace{ 30 pt } 
    When the bosons condense the  gauge fields propagators become massive. Note that the two-legs vertices  in ${\bf a}_\mu$-${\bf a}_\nu$ are protecting the mass sector\cite{note6}. This is shown in Appendix \ref{app:WI}, using again the Ward identity associated to the gauge invariance of the problem. It is  checked as well  in Appendix \ref{app:masses} by a direct evaluation at the first order.

\section{Eliashberg theory}
\label{sec:eliashberg}

 Our QCP  corresponds to a $q=0$ transition.
 In the two previous sections we have treated the fluctuations within the RPA,  for  which the polarization is evaluated at the first loop order. A self consistent treatment of the fluctuations is needed to control the results.
   Integrating the fermions  out  of the partition function is a dangerous uncontrolled step  for such transitions. A better  approach is  the Eliashberg theory, controlled in a large N expansion.   For the details we refer the reader to the extensive review of this technique given in\cite{jerome}.  For completeness we recall here  the  reasoning and the main results.
  The Eliashberg theory relies on three steps.  The first step is to neglect  the renormalization of the vertices as well as the momentum dependence of the self-energy. In the second step, the Dyson  equation is used to evaluate  self-consistently the boson polarization and the fermion's self-energy.  Then one  checks that the   approximation is correct. Here  we have two types  of  fermions and  two type of massless bosons as well - the order parameter and the vector gauge fields; the  time gauge field $a_0 $ being massive.  The coupled Dyson equations are represented below with  Feynman diagrams 
  \bea \label{self1}
  G_f^{-1} ( k , \omega_n )  & =   & i \omega_n - \epsilon_0  + i \Sigma_f ( \omega_n ) ;  \\
  G_c^{-1}  ( k , \omega_n ) & = &  i \omega_n - \epsilon_k  + i \Sigma_c ( \omega_n ) ;  \nonumber \\
  D^{-1}_b ( q, \Omega_n ) & = &  \rho_0 \left [ - i  \Omega_n + \delta + a  q ^2  + \Pi_{fc} ( q, \Omega_n )  - \Sigma_b ( \Omega_n)  \right ]; \nonumber \\
  D^{-1}_{ij}  ( q, \Omega_n) & = &  \rho_0  \left [  (q/k_F)^2  + \Pi_{f} ( q, \Omega_n ) + \Pi_b (q, \Omega_n )  \right  ] \nonumber \\
  & \times&   ( \delta_{ij}  - q_i q_j / q^2 )  \nonumber  \ , \eea where 
  \vspace{ - 50 pt } 
  \begin{center}
  $\Sigma_f ( \omega_n ) = $ \begin{picture}(90,70)(0,28)
\Text(10,20)[c]{$f$}
\Text(35,20)[c]{$c$}
\Text(35,55)[c]{$b, a_\mu$}
\Vertex(50,30){1.5}
\Vertex(20,30){1.5}
\GlueArc(35,30)(15,0,180){2}{6.5}
\LongArrowArc(35,30)(10,60,120)
\ArrowLine(20,30)(50,30)
\DashArrowLine(0,30)(20,30) {1.5}
\DashArrowLine(50,30)(70,30){1.5}
\end{picture}   ; 
\end{center} 
\vspace{ -  35 pt}
  \begin{center}
  $\Sigma_c ( \omega_n ) = $ \begin{picture}(90,70)(0,28)
\Text(10,20)[c]{$c$}
\Text(35,20)[c]{$f$}
\Text(35,55)[c]{$b, a_\mu$}
\Vertex(50,30){1.5}
\Vertex(20,30){1.5}
\GlueArc(35,30)(15,0,180){2}{6.5}
\LongArrowArc(35,30)(10,60,120)
\DashArrowLine(20,30)(50,30){1.5}
\ArrowLine(0,30)(20,30)
\ArrowLine(50,30)(70,30)
\end{picture}  ; 
\end{center} 
\begin{center}
 [\begin{picture}(20,30)(18,0) 
   \Gluon(20,2)(40,2){2}{4} 
  \end{picture} ]
$^{-1} = $ [\begin{picture}(20,30)(18,0) 
   \ZigZag(20,2)(40,2){2}{4} 
  \end{picture} ]
$^{-1}  \ + \ $  [\begin{picture}(20,30)(18,0) 
   \Photon(20,2)(40,2){2}{4} 
  \end{picture} ]
$^{-1} $;
\end{center}
\vspace{ -  35 pt}
  \begin{center}
  $\Sigma_b ( \Omega_n ) = $ \begin{picture}(90,70)(0,28)
\Text(10,20)[c]{$b$}
\Text(35,55)[c]{$ a_\mu$}
\Vertex(50,30){1.5}
\Vertex(20,30){1.5}
\GlueArc(35,30)(15,0,180){2}{6.5}
\LongArrowArc(35,30)(10,60,120)
\Photon(20,30)(50,30){2}{4.5}
\Photon(5,30)(20,30){2}{2.5}
\Photon(50,30)(65,30){2}{2.5}
\end{picture}  ; 
\end{center} 
\vspace{ - 35 pt }
  \begin{center}
  $\Pi_{fc} (q, \Omega_n ) = $  \begin{picture}(90,70)(0,90)
\Text(8,99)[c]{$b$}
\Text(35,113)[c]{$c$}
\Text(35,63)[c]{$f$}
\LongArrow(6,85)(12,85)
\Vertex(20,90){1.5}
\Vertex(50,90){1.5}
\Photon(5,90)(20,90){2}{2.5}
\Photon(50,90)(65,90){2}{2.5}
\DashArrowArcn(35,90)(15,0,180){1.5}
\ArrowArcn(35,90)(15,180,0)
\end{picture}  ; 
\end{center} 
\vspace{ 30 pt } 
\vspace{ -  45 pt}
  \begin{center}
  $\Pi_{f} (q, \Omega_n ) = $  \begin{picture}(90,70)(0,90)
\Text(8,100)[c]{${\bf a}_\mu$}
\Text(59,100)[c]{${\bf a}_\nu$}
\Text(35,115)[c]{$f$}
\Text(35,63)[c]{$f$}
\LongArrow(6,85)(12,85)
\LongArrow(55,85)(60,85)
\Vertex(20,90){1.5}
\Vertex(50,90){1.5}
\ZigZag(5,90)(20,90){2}{2.5}
\ZigZag(50,90)(65,90){2}{2.5}
\DashArrowArcn(35,90)(15,0,180){1.5}
\DashArrowArcn(35,90)(15,180,0){1.5}
\end{picture} ; 
\end{center} 
\vspace{ 30 pt } 
\vspace{ -  45 pt}
  \begin{center}
  $\Pi_{b} (q, \Omega_n ) = $  \begin{picture}(90,70)(0,90)
\Text(8,100)[c]{${\bf a}_\mu$}
\Text(59,100)[c]{${\bf a}_\nu$}
\Text(35,115)[c]{$b$}
\Text(35,63)[c]{$b$}
\LongArrow(6,85)(12,85)
\LongArrow(55,85)(60,85)
\Vertex(20,90){1.5}
\Vertex(50,90){1.5}
\ZigZag(5,90)(20,90){2}{2.5}
\ZigZag(50,90)(65,90){2}{2.5}
\PhotonArc(35,90)(15,0,180){2}{6}
\PhotonArc(35,90)(15,180,0){2}{6}
\LongArrowArc(35,90)(10,60,120)
\end{picture}  ; 
\end{center} 
\vspace{ 30 pt } 
Note that, in the above diagrams, the lines are full by construction, but the evaluation of the diagram doesn't change if the 
fermion propagators are taken to be bare.
The bosonic ``self -energy'' $ \Sigma_b$  and the polarization $ \Pi_b$ are given in Appendix \ref{app:polarization}.
  
   The crucial point in the Eliashberg theory, is to check that the three-legs vertices are small.  We recall that, in the regime $T\geq E^*$,  the dynamical exponent is $z=3$.  This exponent  characterizes as well the ferromagnetic  QCP\cite{jerome}  and the U(1) gauge theories\cite{aim}, which have been studied in the literature. In both cases, the Eliashberg theory is controlled in the same way, using  the combined effects of the large N expansion  and the curvature of the Fermi surface. We recall here the results.
  
   There are two types of vertices depending on the incoming momentum and frequency. 
    The static vertex,
    \vspace{-30 pt}  
  \begin{center}
    $  \Gamma(0,0) \; $= \begin{picture}(70,70)(0,33)
\Text(12,43)[c]{$0,0$}
\Text(63,2)[c]{$k_F,0$}
\Vertex(20,35){1.5}
\Photon(5,35)(20,35){2}{2.5}
\ArrowLine(20,35)(38,47)
\ArrowLine(50,17)(38,23)
\ArrowLine(50,53)(65,62)
\DashArrowLine(65,8)(50,17){1.5}
\DashArrowLine(38,23)(20,35){1.5}
\DashArrowLine(38,47)(50,53){1.5}
\Photon(38,23)(50,53){2}{4.5}
\Photon(50,17)(38,47){2}{4.5}
\Vertex(50,17){1.5}
\Vertex(50,53){1.5}
\Vertex(38,23){1.5}
\Vertex(38,47){1.5}
\end{picture} 
\end{center}
\vspace{30 pt }  obtained for vanishing incoming frequency, is small in the large N  limit, going like $ \Gamma(0,0) \sim N^{-1/2} $.     The dynamical vertex, however, obtained in the limit of non  vanishing, but  still small frequency  when $ |  \Omega_n | \ll v_F q $  is dangerous.  Note that within our model  it is impossible  to form the mixed $fc$-vertex at the first order.  Hence we form it at the second order, but evaluate each loop separately in Appendix \ref{app:vertices}
\vspace{-30 pt}  
  \begin{center}
    $  \Gamma(q,\Omega) \; $=\begin{picture}(70,70)(0,33)
\Text(12,43)[c]{$0,0$}
\Text(63,2)[c]{$k_F,0$}
\Vertex(20,35){1.5}
\Photon(5,35)(20,35){2}{2.5}
\ArrowLine(20,35)(38,47)
\ArrowLine(50,17)(38,23)
\ArrowLine(50,53)(65,62)
\DashArrowLine(65,8)(50,17){1.5}
\DashArrowLine(38,23)(20,35){1.5}
\DashArrowLine(38,47)(50,53){1.5}
\Photon(38,23)(50,53){2}{4.5}
\Photon(50,17)(38,47){2}{4.5}
\Vertex(50,17){1.5}
\Vertex(50,53){1.5}
\Vertex(38,23){1.5}
\Vertex(38,47){1.5}
\end{picture} 
\end{center}
\vspace{30 pt }   For bare fermion legs, it behaves as
     $ \Gamma ( q, \Omega_n ) \sim  \Omega_n^{ (3 - d )  /3} $ (for $d=3$ we get $ \Gamma ( q, \Omega_n ) \sim -  \Omega_n {\rm Log } |\Omega_n| $ ) hence showing strong anomalies in the frequency dependence in the  infra-red sector.
  Fortunately, in the Eliashberg theory,  the fermion legs are not bare, but dressed with their proper self-energy. For $z=3$ the spinon and electron self-energy behaves as $ \Sigma (  \omega_n ) \sim |\omega_n|^{(d-3)/3} $ (with for $d=3$ , $ \Sigma ( \omega_n) \sim  - \omega_n  {\rm Log } |\Omega_n |  $ ).  Dressing the fermion legs  effectively kills the divergence in frequency. A subtle point is that, with a linearized Fermi surface, the resulting vertex is of order one, with no small parameter. Including the curvature of the Fermi surface changes this situation. In the case where $ q_y^2/ m^* \gg v_f q_x$ where $m^*  $ is the curvature mass,  we  finally get  $ \Gamma ( q, \Omega_n ) \sim \beta^2 Log \beta^2 $ with
 $ \beta m^*/ ( N   m )  \ll 1 $.  As we see, the justification of the Eliashberg theory in the case of a $z=3$ , $q=0$ QCP is not simple; it requires both the presence of the curvature and an additional large N expansion (N being here the number of species of fermions),  which is  rather artificial.  An additional caveat is that the large N  {\it limit} cannot be rigorously taken since, if we did this,  we would not include the diagrams corresponding to the fermionic self-energy  which scale like $N^{-1/3}$ in  $D=2$, hence are sub-leading in the large N limit.  The Eliashberg theory thus relies on a large N ``expansion'' rather than on a ``large N limit''. This remark follows the similar observation made in the previous section, that the form of the boson propagator is not homogeneous in N.
 
   Another regime of interest is the case where $z=2$ ,  for intersecting Fermi surfaces. The theory is equivalent to the AF QCP  whose Eliashberg spin fermion treatment is reviewed in\cite{chubAF}. We refer the reader to this paper to get convinced of the smallness of the vertices.
  
     The two regimes (ii) and (iii) of the boson propagator  are not physically relevant  since the boson  propagator has only a small spectral weight  in this regime compared to regime  (iv).  We still must check that the vertices are small.  Let's check the regime (ii) where the frequency part  has a characteristic logarithmic logarithmic dependence.  In both 2 D and 3D we find that the static vertex scales like  $ \Gamma (0, 0)  \sim  1/N $. Curvature is needed to regularized the dynamical vertex,  which  goes like $ \Gamma  ( q, \Omega_n ) \sim  ( \Omega_n {\rm Log } | {\rm Log }  |\Omega_n | | ) /{\rm  Log } | \Omega_n |  $. Hence large N is not needed   for the smallness of the dynamical vertex in this regime.   Details of the vertex calculation  are given in Appendix \ref{app:vertices}. 
     
     \vspace{0.3 cm}
      Last, the reader may wonder whether the analogous of the  singularities discovered by Belitz, Kirkpatrick and Vojta (BKV)\cite{BKV} exist in this theory. This issue lies beyond the  Eliashberg theory, but is of importance for the stability of $q=0$ QCPs. It has been shown by the authors of \cite{BKV} that  singularities appear in the static sector, in the computation of the static, and temperature dependent polarization. Indeed close to a $q=0$ QCP  of ferromagnetic type, the static polarization \vspace{- 40 pt}\begin{center}
  $\Pi (q, T ) = $  \begin{picture}(90,70)(0,90)
\Text(8,99)[c]{$S$}
\Text(35,113)[c]{$c$}
\Text(35,63)[c]{$c$}
\LongArrow(6,85)(12,85)
\Vertex(20,90){1.5}
\Vertex(50,90){1.5}
\Photon(5,90)(20,90){2}{2.5}
\Photon(50,90)(65,90){2}{2.5}
\ArrowArcn(35,90)(15,0,180)
\ArrowArcn(35,90)(15,180,0)
\end{picture}
\end{center}\vspace{ 30 pt}vanishes at the first loop order.  Technically, one can check that, with zero incoming frequencies, poles of the fermion lines are in the same half-plane; hence the static polarization vanishes at one loop. To get a non zero result one must include
     the first vertex correction. 
     
      In our case, it is worthwhile to notice first, that the one loop polarization has a temperature dependence at lower energies, due to the fact that  we have two fermion species.   Indeed we find by direct computation an activated behavior in the case of gaped Fermi surfaces:
     \[ \Pi_{fc}(T) =  N \frac{\rho_0}{(1 - \alpha^\prime) } \ n_F\left [ \frac{ \alpha^\prime v_F q^* }{ ( 1 - \alpha^\prime) }  \right ] \ , \] where
     $n_F$ is the Fermi distribution function.  This activated behavior is quite small. We have as well  a source of damping  coming from the gauge  fields and leading to the temperature dependence  (see Appendix\ref{app:polarization})
     \[ \Sigma_b (T) \sim T^{(d+2)/2} \ . \] However the BKV  singularity comes from  inserting  vertices and self-energy (beyond Eliashberg theory)  in the polarization bubble. It is dangerous because  the first insertion leads to a contribution with {\it negative} sign, hence de-stabilizing the QCP.
     Here we first observe that there is no first order  vertex correction to the polarization $ \Pi_{fc}$; the diagram simply cannot be formed. 
      At the first order one has only self energy insertions of the type
      \vspace{- 40 pt}
      \begin{center} $ \Pi_{fc}^{(1)} (T) = $
       \begin{picture}(90,70)(0,88)
\Text(8,99)[c]{$b$}
\Text(70,99)[c]{$b$}
\Vertex(20,90){1.5}
\Vertex(60,90){1.5}
\Photon(5,90)(20,90){2}{2}
\Photon(60,90)(75,90){2}{2}
\ArrowArc(40,90)(20,0,30)
\DashArrowArc(40,90)(20,30,150){1.5}
\PhotonArc(40,128)(30,-120,-60){2}{4}
\ArrowArc(40,90)(20,150,180)
\DashArrowArc(40,90)(20,180,0){1.5}
\end{picture} $ + $  \begin{picture}(90,70)(0,88)
\Text(8,99)[c]{$b$}
\Text(70,99)[c]{$b$}
\Vertex(20,90){1.5}
\Vertex(60,90){1.5}
\Photon(5,90)(20,90){2}{2}
\Photon(60,90)(75,90){2}{2}
\DashArrowArc(40,90)(20,0,30){1.5}
\ArrowArc(40,90)(20,30,150)
\PhotonArc(40,128)(30,-120,-60){2}{4}
\DashArrowArc(40,90)(20,150,180){1.5}
\ArrowArc(40,90)(20,180,0)
\end{picture} \end{center}\vspace{30 pt}  We find, in the intermediate regime, in $D=3$
\[ \Pi_{fc}^{(1)} \sim - N^{-2/3} T^{4/3} \  . \] We argue that this contribution is smaller than $E^*$ and can be neglected.
In the low energy regime we find that the mismatch between the two Fermi surfaces protects the system, independently of whether the Fermi surfaces intersect or not.

The first vertex correction comes at the two loop order as depicted  below. Note that, contrarily to the nematic case\cite{jerome},  there is no cancellation between the vertices and self energy insertions. This is due to the fact that we have two fermion species, hence the cancellation is spoiled, even though the transition is in the charge channel.  We find that this correction behaves as \[ \Pi_{fc}^{(2)} \sim( N \alpha^\prime)^{- 2/3} T ^{5/3} \] in the intermediate regime in $D=3$, and has a positive sign. It is thus not dangerous.
 \vspace{- 30 pt}
\begin{center} $\Pi_{fc}^{(2)} (T) = $
\begin{picture}(100,70)(-5,88)
\Text(8,99)[c]{$b$}
\Text(90,99)[c]{$b$}
\Vertex(20,90){1.5}
\Vertex(80,90){1.5}
\Vertex(40,110){1.5}
\Vertex(40,70){1.5}
\Vertex(60,110){1.5}
\Vertex(60,70){1.5}
\Photon(5,90)(20,90){2}{2}
\Photon(80,90)(95,90){2}{2}
\ArrowArcn(40,90)(20,180,90)
\Photon(40,110)(60,70){2}{4}
\Photon(60,110)(40,70){2}{4}
\DashArrowLine(40,110)(60,110){1.5}
\ArrowLine(60,70)(40,70)
\DashArrowArcn(40,90)(20,-90,-180){1.5}
\ArrowArcn(60,90)(20,90,0)
\DashArrowArcn(60,90)(20,0,-90){1.5}
\end{picture}  $ \; \; \sim \;   ( N \alpha^\prime )^{-2/3} T^{5/3}  $ \end{center} \vspace{30 pt} Note that  the fermion lines are full in the computation of these diagrams.  The interested reader can find more details about this discussion in Appendix \ref{app:BKV} where we comment as well on $D=2$.

     \section{Ioffe-Larkin (IL) composition rules} 
     \label{sec:IL}    

 In this section we start the study of transport properties of this  quantum critical gauge theory.  A previous study exists\cite{schofield}, but here we re-cast the formalism in terms of the  Ioffe-Larkin composition rules  for  the resistivity and enlarge the discussion to the temperature dependence of the resistivity .
  It is  known from the seminal paper of Ioffe and Larkin\cite{il}, that  in the Coulomb phase of a gauge theory,  both spinons and holons participate to transport and the total  conductivity is constrained by the Ioffe-Larkin (IL) composition rules.   In order to derive them for our model, we must  expose the system to an external electro-magnetic field ${\bf A } $.   ${\bf A } $ is attached to the  conduction electrons and to the holons $b$ \cite{note4}.    The system is as well  subjected  to the internal fictitious gauge field ${\bf a} $ (\ref{eqn8}). ${\bf a} $  is attached to the  fields carrying the  gauge charge, hence to  the holon  $  b $ an the spinon  $ f_ \sigma $.
   We work in the Coulomb gauge, ensuring that  $\nabla \cdot ( {\bf A} + {\bf a} )  = 0 $. The system is invariant through:
\beq \begin{array}{ll}
{\bf A} \rightarrow {\bf A} + \nabla \theta_A \; \;  \ , &  {\bf a} \rightarrow {\bf a} + \nabla \theta_a  \; \ , \\
\\
c \rightarrow c \  {\rm Exp } ( i \theta_A )  \; \; \, & f \rightarrow f \  {\rm Exp } ( i \theta_a )  \; \ , \\
\\
b \rightarrow b  \ {\rm Exp } ( i  \theta_A - i \theta_a  )   \  .  & \end{array} \eeq  In order to derive the IL composition rules, we expand the action for minimal coupling with the vector gauge fields.   Note that the composition rules are completely general (see Appendix \ref{app:IL});  the expansion in  minimal coupling
is a matter of convenience.
\bea \label{eqn42}
S& = &  \int   d \tau \sum_k  f^\dagger G_f^{-1} ( a_\mu) f   \\
& + & c^\dagger G_c^{-1}( A_\mu ) c + b^\dagger D_b^{-1} (A_\mu, a_\mu ) b ; \nonumber \\
 & \simeq &\int d \tau \sum_k \left [  f^\dagger \ G_{0,f}^{-1} \ f  + c^\dagger \ G_{0,c}^{-1} \ c + b^\dagger \ D_{0,b}^{-1} \ b  \right .  \nonumber \\
 & + & \left .  f^\dagger \  {\bf v}_f \cdot {\bf a }  \   f +   c^\dagger  \  {\bf v}_c \cdot {\bf A } \  c +  b^\dagger \  {\bf v}_b \cdot  ({\bf a } + {\bf A} )  \  b \right ]  , \nonumber \eea
 where the vertices ${\bf v}_f $, ${\bf v}_c $, ${\bf v}_b $ are determined from the expansion   of the effective lagrangian to the first order in ${\bf a}$, ${\bf A }$\cite{note8}. 
  After integrating out the matter fields $f$,$c$ and $b$, the effective action for minimal coupling to the gauge fields reads:
\bea \label{eqn17} 
S\{A, a\}  & = &  \frac{T}{2} \int \frac{ d^d k }{( 2 \pi)^d}\sum_{\omega_n} \left [ {\bf A}_\alpha (\omega, k )  \Pi_c^{\alpha \beta}  (\omega, k ) {\bf A}_\beta
\right . \nonumber \\
& + & ( {\bf A}_\alpha -{\bf a}_\alpha ) \Pi_b^{\alpha \beta}  (\omega, k ) ( {\bf A}_\beta -{\bf a}_\beta )   \nonumber \\
  & + & \left .   {\bf a}_\alpha (\omega, k )  \Pi_f^{\alpha \beta}  (\omega, k ) {\bf a}_\beta \right ]   \ . \eea  The polarization bubble $\Pi (\omega, k)$  can be decomposed into its longitudinal and transverse parts:
\beq \Pi_{\alpha \beta} = \left ( \delta_{\alpha \beta}  - \frac{ k_\alpha k _\beta} {k^2} \right ) \Pi_1
 + \frac{ k_\alpha k _\beta} {k^2}  \Pi_2 \ . \eeq Following IL, we integrate over the  fictitious gauge field $ { \bf a} $ in (\ref{eqn1})
 to get the effective action for the external electric field.
\beq
S\{A, a\} = \frac{T}{2} \int d k \sum_\omega  {\bf A}_\alpha (\omega, k )
 \left [  \Pi_c + ( \Pi_b^{-1} + \Pi_f^{-1} )^{-1} \right ]
  {\bf A}_\beta  \ . \eeq  Hence the total polarizability of the system is  \beq \label{eqn18}
P_1 = \Pi_c + \left ( \Pi_f^{-1} + \Pi_b^{-1} \right )^{-1} \ . \eeq
  Using the Kubo formula
  \beq \sigma = \frac{i \Pi(\omega  ) }{ \omega} \ , \eeq we get for the conductivity 
\beq \label{eqn19} 
\sigma = \sigma_c + ( \sigma_f^{-1} + \sigma_b^{-1})^{-1} \ . \eeq Note that the composition rule for the conductivity can be derived in a simpler way,  using solely the  constraint attached to the gauge symmetry; this calculation is given in Appendix \ref{app:IL}.
    The composition rules (\ref{eqn18}) are powerful enough to allow  a complete discussion of  the  electrical transport close to the  QCP. 
    One observes that the holons are ``sandwiched'' between the spinons  and the conduction electrons.  If $\Pi_b^{-1}$ dominates $\Pi_f^{-1}$ , then it is very unlikely that  $\Pi_b$ will dominate $\Pi_c$ since  $\Pi_c $  already dominates $\Pi_f$. Hence we  infer  from the simple form of (\ref{eqn18}) that  $\Pi_c$ is the most important part. The transport is dominated  the conduction electrons.
        
    Let's first examine the  limit of zero temperature.    On the heavy Fermi liquid side, the holons  $b$  are perfect conductors and the spinons are  massless fermions so that
    \beq \begin{array}{ll}
    {\displaystyle \Pi_b = \frac{n_b e^2}{m_b} } ; \;  \;  & {\displaystyle  \Pi_f = \frac{n_f e^2}{m_f} \ \frac{  - i \omega \tau_f }{ ( 1 - i \omega \tau_f  ) } }, \end{array}  \eeq where  we have used a Drude formula\cite{mahan} for the polarization of the f-spinons  and $\tau_f$ is the scattering lifetime of the spinons\cite{note5}, which writes $\tau_f^{-1} \sim \tau_0^{-1} + T^2 $ in the  heavy Fermi liquid. Note that we have taken into account the effect of impurities in the  scattering time of  the spinons. This corresponds to dressing the spinon lines with disorder, but neglecting the vertex corrections. In the limit of low frequencies, $ \Pi_b^{-1} + \Pi_f^{-1} $ is fully dominated by  the second term;  hence the holons don't affect the residual conductivity. 
  We get, on the heavy Fermi liquid side and in the limit of low frequencies,
  \beq \label{eqn20}
  \Pi_1 \simeq  \Pi_c + \Pi_f \ , \eeq  \beq \label{eqn23} \mbox{with} \; \; \;  \Pi_c = \frac{n_c e^2}{m_c} \ \frac{  - i \omega \tau_c }{ ( 1 - i \omega \tau_c  ) }  \ . 
  \eeq  In (\ref{eqn20}) the effect of impurities is implicitly taken into account in the scattering lifetime of the conduction electrons  $\tau_c^{-1} \sim \tau_0^{-1} + T^2 $. In the limit of zero frequencies we get
  \beq  \label{eqn33} \sigma_1 = \frac{ n_c e^2 \tau_c}{m_c}  + \frac{ n_f e^2 \tau_f}{m_f}  \  , \eeq while, on the   localized 
  side of the transition,  only the conduction electron conduct. The residual conductivity thus jumps at the transition;  on the heavy Fermi liquid side the f-spinon start  abruptly to conduct. This result is in agreement with the study of\cite{schofield}.  Note  however, that the jump in the conductivity obtained in this model  is very unlikely to be detectable , since $m_c \ll m_f $\cite{note5}.
  
    We turn next to  the temperature dependence of the conductivity at the QCP.  We first focus on the low temperature regime, already studied in\cite{ssv}, saving the study of the intermediate temperature regime for the next section.
      As  in the previous section, there are two cases of  interest.  In case a) there is a gap in the continuum of the  particle-hole excitations.   Because of the  gap, the  electron and spinon lifetime are not affected by the scattering with the holons, and we get  the standard Landau Fermi liquid law $\tau_f^{-1} \sim \tau_c^{-1} \sim  \tau_0^{-1} +  T^2 $.  The damping of the holons doesn't come from the  particle-hole continuum but only from the gauge fields $ \Sigma_b$. This damping of the holons itself produces a finite  bosonic lifetime.  The polarization writes (see Appendix \ref{app:polarization}) \beq
     \Pi_ b = \frac{e^2}{m_b} \left (  - i \omega \tau_b  \right ) \  ,  \eeq 
     \beq \mbox{with \hspace{0.5 cm} }  \begin{array}{ll} 
     \tau_b \sim  - {\rm Log } \ T \;  & \mbox{in } \; D = 2  ; \\
\tau_b \sim  T ^{5/4} \;  &  \mbox{in} \; D = 3; \end{array} \eeq 
     One sees that, in that case as well, the sum $\Pi_f^{-1}+\Pi_b^{-1}$ is dominated by $\Pi_b^{-1}$, for  $D=2$ and by $ \Pi_f^{-1}$ for $ D=3$.    Nevertheless,  in both cases the conductivity is dominated by the conduction electrons as can been seen from (\ref{eqn18}).
       The resistivity thus varies like $T^2$ at the QCP.
       \beq
\rho \sim  \rho_0 +  T^{2} \; \mbox{in} \;  D=3 \  . \eeq  This result contradicts the previous study \cite{ssv}.  In our view the contribution of the conduction electrons to the  conductivity was overlooked in \cite{ssv}. 
    
    The second case of interest is when the two Fermi surfaces intersect, called case b).  Here the particle hole continuum  has no gap, hence the bosons are not  perfect conductors anymore, but are damped by the particle-hole continuum.  We write (see Appendix \ref{app:polarization} )  \beq
    \label{eqn22} \Pi_ b = \frac{e^2}{m_b} \left (  - i \omega \tau_b  \right ) \  , 
  \eeq   where $\tau_b$ carries the temperature dependence of the polarization.  We find for $z=2$
  \beq \begin{array}{ll}    
\tau_b \sim - {\rm Log} \  T   \;  & \mbox{in } \; D = 2  ; \\
\tau_b \sim \sqrt{T} \;  &  \mbox{in} \; D = 3; \end{array} \eeq 
Now in the Eliashberg theory,  damping from the gauge fields leads to, $\tau_f^{-1} \sim \tau_0^{-1}  + \Sigma_f (T)$ with \[ \Sigma_f (T) \sim \left \{ \begin{array}{ll}
 T^{2/3}  \;  &  \mbox{for} \; D = 2 ; \\
 - T {\rm Log } \ T   \;  & \mbox{for } \; D = 3   \  . \end{array} \right .  \] 
One sees that, in that case as well, the sum $\Pi_f^{-1}+\Pi_b^{-1}$ is dominated by $\Pi_b^{-1}$, for  $D=2$ and by $ \Pi_f^{-1}$ for $ D=3$.    Nevertheless,  in both cases the conductivity is dominated by the conduction electrons. Since  the regime is  characterized by $z=2$, the  conduction electrons have the inverse scattering lifetime  $\Sigma_c \simeq T^{3/2} $ in $D=3$.   Since  the  backscattering processes are naturally present in the model (see next section), this regime is equivalent to the AF SDW  scenario, with
\beq
\rho \sim  \rho_0 +  T^{3/2} \; \mbox{in} \;  D=3 \  . \eeq 

 A last remark for  this section is that we recover  the result of  \cite{schofield} that, although the  gauge fields are gapped in the Higgs phase,  the external electromagnetic field is not, which  prohibits -thank goodness ! -Meissner effect in the heavy Fermi liquid phase. The easiest way to see it is to make the change of variable $ a^\prime_\mu = a_\mu + A_\mu $ in (\ref{eqn42}) .   The fictitious gauge fields are now $a^\prime_\mu$ and the external gauge fields are decoupled  from the holons :
 \beq
 S_{int} =   b^\dagger \  {\bf v}_b \cdot {\bf a }^\prime  \   b +   c^\dagger  \  {\bf v}_c \cdot {\bf A } \  c +  f^\dagger \  {\bf v}_f \cdot  ({\bf a }^\prime - {\bf A} )  \  f  \  . \eeq  One now follows Appendix \ref{app:WI} and write the Ward Identity related to the {\it external} gauge field. Since the source terms can  be put to zero (  only the holons get a non vanishing source term) the mass of the electromagnetic field  vanishes like
 \beq - i q^\mu \Pi_{A_\mu A_\nu } = 0  \ . \eeq

\section{ Quasi-linear resistivity  } 
\label{sec:resistivity}  

 We examine  here the transport in the intermediate regime.  This regime differs from the low temperature one by its dynamical exponent which is now $z=3$. In this regime, the IL composition rules  are still valid, with 
  \beq \begin{array}{ll}    
\tau_b \sim - {\rm Log  } \  T    \;  & \mbox{for } \; D = 2  ; \\
\tau_b \sim  T^{1/3} \;  &  \mbox{for} \; D = 3; \end{array} \eeq   hence the  holons still don't participate to the transport. As before, the conductivity is dominated by the conduction electrons. In this section we examine in more details the Drude form assumed in (\ref{eqn20}) .  The arguments have been given in a previous work, and we recall them here for clarity\cite{us}.
   The main idea is that the Drude form is valid, with  the scattering lifetime of the conduction electrons given by
   \beq \label{eqn24}
   \tau^{-1}_c \sim -T {\rm Log } \  T   \eeq in $D=3$.  (\ref{eqn24})  is the typical scattering lifetime for  a $z=3$  QCP, like for example a ferromagnet.    The important question is  
   --``how does the current decay in this special $ q =0$ phase transition ?''  
    The unique feature of this QCP is that the current naturally decays through the lattice of f- electrons. Contrarily to a usual ferromagnetic QCP or  nematic QCP where translational invariance is not broken at the phase transition, here  translational invariance is naturally broken since the f-electrons  are on the brink of localization. Hence, there is no need for external translation invariance breakers, like impurities, to break translational invariance.  Umklapp processes are naturally present  which decay the current.
    In this sense
    \beq
    \tau_{tr} \simeq \tau_{QP} \  ,\eeq where $\tau_{tr}$ is the transport time while $\tau_{QP}$ is the quasi-particle lifetime. For $z=3$,  $D=3$,  it has the standard form
    \beq
    \tau_{QP}^{-1} \sim T  \  . \eeq  Our claim is that the electric transport in this phase is correctly described by  the  Drude polarization
    \vspace{- 30 pt}
    \begin{center}
   $ \Pi_c  $  =\begin{picture}(90,70)(0,88)
\Text(8,99)[c]{$J_c$}
\Text(70,99)[c]{$J_c$}
\Vertex(20,90){1.5}
\Vertex(60,90){1.5}
\ArrowLine(5,90)(20,90)
\ArrowLine(60,90)(75,90)
\ArrowArc(40,90)(20,0,180)
\ArrowArc(40,90)(20,180,0)
\end{picture} $ + $   \begin{picture}(90,70)(0,88)
\Text(8,99)[c]{$J_c$}
\Text(70,99)[c]{$J_c$}
\Vertex(20,90){1.5}
\Vertex(60,90){1.5}
\ArrowLine(5,90)(20,90)
\ArrowLine(60,90)(75,90)
\ArrowArc(40,90)(20,0,30)
\DashArrowArc(40,90)(20,30,150){1.5}
\PhotonArc(40,128)(30,-120,-60){2}{4}
\ArrowArc(40,90)(20,150,180)
\ArrowArc(40,90)(20,180,0)
\end{picture}
\end {center}
\vspace{ 30 pt }  , where here we have used bare lines. 
    More precisely, using the Kubo formalism, one can first show that the  conduction electron polarization is unaffected by coupling to  the  bosons and the f-fermions.  The coupling to f and b  is protected by gauge invariance. Pictorially we mean that
    \vspace{ - 30 pt }
  \begin{center}
    \begin{picture}(90,70)(0,88)
\Text(8,99)[c]{$J_c$}
\Text(70,99)[c]{$J_f$}
\Vertex(20,90){1.5}
\Vertex(60,90){1.5}
\ArrowLine(5,90)(20,90)
\ArrowLine(60,90)(75,90)
\ArrowArcn(40,90)(20,180,90)
\Photon(40,110)(40,70){2}{4}
\ArrowArcn(40,90)(20,-90,-180)
\DashArrowArcn(40,90)(20,90,0){1.5}
\DashArrowArcn(40,90)(20,0,-90){1.5}
\end{picture} $ + $ \begin{picture}(90,70)(0,88)
\Text(8,99)[c]{$J_c$}
\Text(70,99)[c]{$J_b$}
\Vertex(20,90){1.5}
\Vertex(60,90){1.5}
\ArrowLine(5,90)(20,90)
\ArrowLine(60,90)(75,90)
\ArrowArcn(40,90)(20,180,90)
\DashArrowLine(40,110)(40,70){1.5}
\ArrowArcn(40,90)(20,-90,-180)
\PhotonArc(40,90)(20,0,90){2}{3}
\PhotonArc(40,90)(20,-90,0){2}{3}
\end{picture} $ + $
\end{center}
\vspace{ - 30 pt }
\begin{center}
\begin{picture}(150,90)(0,88)
\Text(8,99)[c]{$J_c$}
\Text(130,99)[c]{$J_b$}
\ArrowLine(5,90)(20,90)
\ArrowLine(120,90)(135,90)
\Vertex(20,90){1.5}
\Vertex(60,90){1.5}
\Vertex(80,90){1.5}
\Vertex(120,90){1.5}
\ZigZag(60,90)(80,90){2}{4}
\ArrowArcn(40,90)(20,180,90)
\DashArrowLine(40,110)(40,70){1.5}
\ArrowArcn(40,90)(20,-90,-180)
\PhotonArc(40,90)(20,0,90){2}{3}
\PhotonArc(40,90)(20,-90,0){2}{3}
\PhotonArc(100,90)(20,0,180){2}{7}
\PhotonArc(100,90)(20,180,0){2}{7}
\end{picture} $ + $
\end{center}
\vspace{- 30 pt}
\begin{center}
\begin{picture}(150,90)(0,88)
\Text(8,99)[c]{$J_c$}
\Text(130,99)[c]{$J_f$}
\ArrowLine(5,90)(20,90)
\ArrowLine(120,90)(135,90)
\Vertex(20,90){1.5}
\Vertex(60,90){1.5}
\Vertex(80,90){1.5}
\Vertex(120,90){1.5}
\ZigZag(60,90)(80,90){2}{4}
\ArrowArcn(40,90)(20,180,90)
\Photon(40,110)(40,70){2}{4}
\ArrowArcn(40,90)(20,-90,-180)
\DashArrowArcn(40,90)(20,90,0){1.5}
\DashArrowArcn(40,90)(20,0,-90){1.5}
\DashArrowArcn(100,90)(20,180,0){1.5}
\DashArrowArcn(100,90)(20,0,-180){1.5}
\end{picture}  $ = 0 $
\end{center}\vspace{30 pt }  This comes from the fact that 
  [\begin{picture}(20,30)(18,0) 
   \ZigZag(20,2)(40,2){2}{4} 
  \end{picture} ]
$^{-1} = \Pi_f  + \Pi_b$. 
The next observation is that the vertex corrections are negligible in this regime; namely the next leading order to the Drude formulae, is of order  $ (\alpha^\prime)^2$ and  $1/N$, as shown in the diagram below.
\vspace{- 30 pt}
\begin{center}
\begin{picture}(150,70)(-50,88)
\Text(8,99)[c]{$J_c$}
\Text(90,99)[c]{$J_c$}
\Vertex(20,90){1.5}
\Vertex(80,90){1.5}
\Vertex(40,110){1.5}
\Vertex(40,70){1.5}
\Vertex(60,110){1.5}
\Vertex(60,70){1.5}
\ArrowLine(5,90)(20,90)
\ArrowLine(80,90)(95,90)
\ArrowArcn(40,90)(20,180,90)
\Photon(40,110)(40,70){2}{4}
\Photon(60,110)(60,70){2}{4}
\DashArrowLine(40,110)(60,110){1.5}
\DashArrowLine(60,70)(40,70){1.5}
\ArrowArcn(40,90)(20,-90,-180)
\ArrowArcn(60,90)(20,90,0)
\ArrowArcn(60,90)(20,0,-90)
\end{picture}  $ \simeq \;   [ (\alpha^\prime)^2 / N ]  $ \end{center} \vspace{30 pt}
   Hence, in this regime the transport is simple and electrical conductivity can be expressed through  the simple Drude formulae.
   The evaluation of the inverse scattering time in $D=3$ is given in Appendix\ref{app:imsigma}.    We get \beq \begin{array}{l}
    \tau_{tr} \sim T {\rm Log }  (T/E^*)  \; \mbox{in} \; D=3 \\
    \tau_{tr} \sim T^{2/3}  \; \mbox{in} \; D=2 \ . \end{array} \eeq
    Note that the logarithm in $D=3$ has a thermal origin; it differs from the logarithm which appears in the  real part of the self-energy, in $D=3$.

\section{Summary of thermodynamic and transport}
\label{sec:summary}

   To summarize transport and thermodynamics close to the QCP, we distinguish the regime I for $T\leq E^*$ and the regime II for $T \geq E^*$.   The exponents for transport and thermodynamics in regime I depend on the form of the spinon Fermi surface.  If there is a gap between the spinon and electron Fermi surfaces, as is show in Fig.\ref{Fermisurfaces} for the  case a), then the  anomalous exponents for the effective mass - appearing in $C_v$, are due to the massless gauge fields  with $z=3$.  The electrical resistivity is dominated  by the conduction electrons, and since the scattering with the  spinons is gapped, it follows the usual $T^2$ law characteristic of the Landau Fermi liquid.   The susceptibility  doesn't couple  directly to the critical modes, hence here again the Fermi liquid law is recovered.   The exponents are summarized in  table I.
\vskip0.5 cm
\begin{table}[h]
\label{table1}\begin{tabular}{c||c|c|c}
 $ T \leq E^* $ Regime I & $ C_v $ & $\Delta \rho (T) $ & $ \chi(T) $\\ \hline
& & & \\
$D=3$ & $- T  {\rm Log } (T)  $ & $ T^2 $& $\chi_0$ \\ \hline
& & & \\
$D=2 $ & $ T^{2/3} $ & $ T^2$ & $ \chi_0 $  \\
\end{tabular} \caption{Transport and thermodynamic exponents in the low temperature regime when the  particle-hole continuum is gapped.}
\end{table}
\vskip 0.5 cm
   We see that the situation is peculiar in the sense that the Landau Fermi liquid paradigm is broken  for the thermodynamics only. The resistivity behaves as $T^2$  even though the residual resistivity jumps at the Fermi surface.   There is no trace of  anomalous exponents   for the dynamic spin susceptibility, fact which poorly fits the experimental observations.
   
    The second case is when $T\leq E^*$, but the  spinon and electron Fermi surfaces intersect. We call this regime I$^\prime$.
    In that case the particle-hole continuum goes down to $T=0$ with  hot regions at the intersection of the two Fermi surfaces.  The situation is analogous to the  AF SDW QCP, except that the critical modes are solely in the charge channel. The results are summarized in table II.
    \vskip0.5 cm
\begin{table}[h]
\label{table2}\begin{tabular}{c||c|c|c}
 $ T \leq E^* $ Regime I$^\prime$ & $ C_v $ & $\Delta \rho (T) $ & $ \chi(T) $\\ \hline
& & & \\
$D=3$ & $- T  {\rm Log } (T) $ & $ T^{3/2} $& $\chi_0$ \\ \hline
& & & \\
$D=2 $ & $T^{2/3} $ & $ -T {\rm Log } T  $ & $ \chi_0 $  \\
\end{tabular} \caption{Transport and thermodynamic exponents in the low temperature regime, where the spinon and electron Fermi surfaces intersect. }
\end{table}
\vskip 0.5 cm

  In our  view, the most interesting  regime is  for $T\geq E^*$. In that case, the exponents  don't  depend on the shape of the spinon Fermi surface, but the very existence of this regime requires the presence of a spinon Fermi surface at the QCP.  Here the transport is simpler than closer to the QCP.   The resistivity shows a quasi linear behavior in $D=3$ and a sub-linear exponent in  $D=2$. The results are summarized in table III.
 \vskip0.5 cm
\begin{table}[h]
\label{table3}\begin{tabular}{c||c|c|c}
 $ T \gg E^* $ Regime II & $ C_v $ & $\Delta \rho (T) $ & $ \chi(T) $\\ \hline
& & & \\
$D=3$ & $ T {\rm Log } (T/E^*) $ & $T {\rm Log} (T/E^*) $& $ T^{4/3}$ \\ \hline
& & & \\
$D=2 $ & $ T^{2/3} $ & $ T^{2/3}$ & $ -T{\rm  Log } (T) $  \\
\end{tabular} \caption{Transport and thermodynamic exponents in the Maginal Fermi liquid regime around the KB-QCP. 
The exponents are in agreement with those of Ref.\cite{us}.}
\end{table}
\vskip 0.5 cm  Note that the temperature dependence of the spin susceptibility, although departing from the standard  Landau Fermi liquid law, still doesn't show anomalous exponents.

 \section{Conclusions}
 \label{sec:conclusions}
 
  In this paper we have given the simplest  consistent treatment of  a selective Mott transition  in the Anderson lattice, using a U(1) slave boson technique associated with an Eliashberg treatment of the vertices.  We find that the QCP exists and  has a multi-scale character.   At the QCP below $E^*$, the exponents for thermodynamics and transport depend on the shape of the  spinon and electron Fermi surfaces. 
   When the particle-hole continuum  is gapped, the   anomalous transport scattering is  gapped as well, and the resistivity follows the  landau Fermi liquid law. The effective mass is dominated by the fluctuations of the transverse gauge fields showing a Reizer singularity\cite{reizer}. If the spinon and electron Fermi surfaces intersect, the particle-hole continuum is gapless and  the transport and thermodynamics are anomalous down to the lowest temperature, with $z=2$ for $T\leq E^*$.   Above the energy scale $E^*$, the resistivity doesn't depend on the shape of the spinon Fermi surface, and we get a universal quasi-linear resistivity  in $D=3$.
     In our view, the important question raised by our study is  the question of  the presence or not of a selective Mott transition in the Anderson lattice. Said in general words, one can reasonably ask whether we believe the anomalous properties 
      observed in heavy fermions are due to a Mott localization of the f-electrons. If  we believe it is the right answer, then the U(1) slave boson gauge theory is the most straightforward approach to such a phenomenon. It would be  very interesting to have studies from other techniques, like DMFT, to first confirm the presence of the transition and, if it there, to give more details  about  the elementary excitations.
    
    It is reasonable to ask whether the U(1) slave boson theory, although  being the simplest description of the Mott transition, is the  appropriated tool.  This question is of relevance as well for  the cuprate superconductors, where gauge theories, with  sometimes bigger algebra  like SU(2), have been extensively used  to  describe the  approach to the Mott state\cite{lee-review,leenagaosa}.  It is clear from the above study, that this approach suffers from what we would call `` spinology''.   Fermionic spinons with a finite Fermi surface are needed at the QCP for the QCP to exist at all.  One can reasonably question this feature, and wonder whether under more powerful techniques, this feature would survive. Nevertheless, before discarding  the U(1) slave boson gauge theory for the Anderson lattice, one must consider that it gives a very interesting regime with a quasi-linear scattering  and transport lifetime in $3 D$. This  unique feature is not easily obtained within any theory and the good point is that it doesn't depend on the shape of the spinon Fermi surface, but  only on its presence at the QCP-hence it is a direct consequence of the ``fractionalization'' of the heavy electron at the QCP. We can use this regime for  making experimentally testable predictions. A  first application to He$^3$ bi-layers has been performed\cite{adel}. We can as well make predictions for the thermal transport in this regime, and call for experimental confirmation  or invalidation.

\vspace{0.5 cm}
 I am indebted  to M. Norman and I. Paul for countless discussions on  the similar Kondo-Heisenberg model.
 Many ideas of this paper came from  interactions with them, related to this previous study.   
The description of  the quasi-linear resistivity benefitted from  a useful discussion with D.L. Maslov and the study 
of the Ward Identities and the  Ioffe-Larkin composition rule, from useful interactions with K-S.  Kim and O. Parcollet,
 and a very useful, heated discussion with P. Coleman. A special thank to C. Bena and A.V . Chubukov for a useful reading of the manuscript.
This work is supported by the French National Grant ANR36ECCEZZZ.

\newpage
\appendix
\section{ Evaluation of the integrals for the mean-field}
\label{app:MF}

   At $T=0$, the calculation of the integrals used in the mean-field treatment is analytical for linearized bands:
  We call 
  \bea  \label{appendeq1}
  {\cal A}  & = &   T \sum _{k, \sigma,  \omega_n }  G_{ff} (k , i \omega_n)  \nonumber  \\
  {\cal B } & = &  T \sum _{k \sigma, \omega_n } G_{fc}   ( k , i \omega_n )   / ( b V + \sigma_0 )  \nonumber \\
  {\cal C} & = &  T \sum_{k, \sigma, \omega_n } \epsilon_k  G_{ff}( k, i \omega_n ) \  .   \nonumber \eea
   We diagonalize the $2x2$ matrix \cite{millis-lee} which accounts for the hybridization of the f- and c- bands:
   \bea 
   X_1 & = &  \frac{1}{2} \left [ \epsilon^0_k + \epsilon_k - \sqrt{\Delta} \right ]\nonumber \\
   X_2 & = &  \frac{1}{2} \left [ \epsilon^0_k + \epsilon_k + \sqrt{\Delta} \right ] \nonumber \\
   \Delta & = & ( \epsilon^0_k  - \epsilon_k ) ^2 + 4 ( \sigma_0 + b V ) ^2  \  . \nonumber \eea
   The integrals are all performed in the same way, first by summing over the Matsubara frequencies , and second by  doing the momentum integration.  The momentum integration is done  by linearization of  the band.
   \bea \label{appendeq2}
   {\cal A} & = &  N T \sum _{k, \omega_n} \frac{ ( i \omega_n - \epsilon_k ) }{  ( i \omega_n -X_1 ) ( i \omega_n - X_2 ) }  \nonumber \\
     & = & N \rho_0 \int_{-D}^{D} d \epsilon  \int \frac{ - n_F (z) } { 2 i  \pi } \ \frac{ (z - \epsilon) }{(z - X_1) (z- X_2)}  \ d z \ , \nonumber \\ 
     & &\mbox{ where the contour is on the whole complex plane } \nonumber \\
     & = &   N \rho_0 \int_{-D}^{D} d \epsilon \left ( \frac{n_F ( X_1)  ( X_1 - \epsilon _k ) }{ ( X_1 - X_2 ) }  - \frac{n_F(X_2) ( X_2 - \epsilon_k) }{( X_1- X_2 )} \right )  \nonumber \\
     & = &  \frac{ N \rho_0}{2} \int_{-D}^{\epsilon_m} d \epsilon \frac{ - y + \sqrt{ y^2 + 4 ( b V + \sigma_0 )^2 }} {  \sqrt{ y^2 + 4 ( b V + \sigma_0 )^2 }}  \nonumber \\
     & - &  \frac{ N \rho_0}{2} \int_{-D}^{\epsilon_p} d \epsilon \frac{ - y - \sqrt{ y^2 + 4 ( b V + \sigma_0 )^2 }} {  \sqrt{ y^2 + 4 ( b V + \sigma_0 )^2 }} \nonumber \  , 
     \eea  with  $ \epsilon_m $ and $\epsilon_p $ the    Fermi levels for the upper and lower bands respectively.
     \bea \label{appendeq3}
     \epsilon_m & = &   (- \epsilon _f + \alpha^\prime \mu  - \sqrt{ ( \epsilon_f + \alpha^\prime \mu )^2 + 4 \alpha ^\prime  ( b V +\sigma_0 ) ^2 }  )   / ( 2 \alpha^\prime )  \nonumber \\
       \epsilon_p & = &   ( - \epsilon _f + \alpha^\prime \mu  + \sqrt{ ( \epsilon_f + \alpha^\prime \mu )^2 + 4 \alpha ^\prime  ( b V +\sigma_0 ) ^2 }  ) /  ( 2 \alpha^\prime  )   \nonumber \\
       && \mbox{ with the conditions } \begin{array}{ll} -D \leq \epsilon_m \leq  0 ; \; \; & 0 \leq \epsilon_p \leq D  \ , \end{array}  \nonumber \\
       && \mbox{and} \;  \alpha^\prime =  \alpha b^2 + \phi_0/ D \ . \nonumber \eea
       One obtains
       \bea {\cal A } &  =  &  \frac{ N \rho_0}{ 2 ( 1 - \alpha^\prime) } \left ( - 2 y_{-D} + y_m - \sqrt{ y_m^2 + 4 ( b V + \sigma_0)^2} \right . \nonumber \\
       &  + & \left .  y _p + \sqrt{ y_p^2 + 4 ( b V + \sigma_0 )^2}  \right )  \  ,  \nonumber \eea
      \[ \begin{array}{ll} \mbox{ where } \; & \begin{array}{l} 
       y_m   =   ( 1 - \alpha^\prime  ) \epsilon_m - \epsilon_f - \mu ;  \\
        y_p   =   ( 1 - \alpha^\prime  ) \epsilon_p - \epsilon_f - \mu ;  \\
         y_{-D}   =   - ( 1 - \alpha^\prime  ) D- \epsilon_f - \mu \  . \end{array} \end{array} \]
         
 We proceed in the same way for ${\cal B } $ and ${\cal C} $  to find:
 \[
 {\cal B } = \frac{ N \rho_0} { ( 1 - \alpha^\prime) }  Log \left [ \frac{ y_m + \sqrt{ y_m^2 + 4 ( b V + \sigma_0 )^2}}{y_p + \sqrt{ y_p^2 + 4 ( b V + \sigma_0 )^2} }\right ] \ . \]
 
 \bea 
{\cal C }   & = &  \frac{ N \rho _0}{ 2 ( 1 - \alpha^\prime ) ^2 } \left [  - 2 ( \epsilon_f + \mu ) ( y _{-D} )  + y_{-D}^2 \right . \nonumber \\
  & +  & 2  (b V + \sigma_0)^2 Log \left (  \frac{ y_m + \sqrt{ y_m^2 + 4 ( b V + \sigma_0 )^2}}{y_p + \sqrt{ y_p^2 + 4 ( b V + \sigma_0 )^2} }\right )   \nonumber \\
 & + &  ( \epsilon_f + \mu ) y_m + y_m^2 \nonumber \\
 & - & ( y_m/2 + \epsilon_f + \mu ) \sqrt{ y _m ^2 + 4 ( b V + \sigma_0 )^2 }  \nonumber \\
  & + &   ( \epsilon_f + \mu ) y_p + y_p^2 \nonumber \\
  &  - & \left .  ( y_p/2 + \epsilon_f + \mu ) \sqrt{ y _p ^2 + 4 ( b V + \sigma_0 )^2 } \right ] \  . \nonumber \eea
  
  \section{Evaluation of the integrals  for  the vertices }
  \label{app:vertices}
  
    In this  section, we focus on the evaluation of the vertices for the regime (ii) where we have logarithmic frequency dependence of the  polarization.  We call  ${\bar g } = V^2 \rho_0 $  with  V  the coupling constant between the soft modes and the f- and c-electrons and $\rho_0$ the density of states of the c-electrons. Note that although it is not possible to form the  vertex correction at the fist loop, we still evaluate the fictitious one-loop vertex below, knowing that the  two-loops vertices behave as products of the one-loop ones.
    We start with the static vertex:
  \vspace{-30 pt}  
  \begin{center}
    $  \Gamma(0,0) \; $= \begin{picture}(70,70)(0,33)
\Text(12,43)[c]{$0,0$}
\Text(63,2)[c]{$k_F,0$}
\Vertex(20,35){1.5}
\Photon(5,35)(20,35){2}{2.5}
\DashArrowLine(20,35)(50,53){1.5}
\ArrowLine(50,17)(20,35)
\ArrowLine(50,53)(65,62)
\DashArrowLine(65,8)(50,17){1.5}
\Photon(50,17)(50,53){2}{4.5}
\Vertex(50,17){1.5}
\Vertex(50,53){1.5}
\end{picture} 
\end{center}
\vspace{30 pt } 
    \bea 
 &&   \Gamma(0,0)  =   \frac{\bar g }{-  N  Log( \alpha ^\prime )} \int \frac{d^d q d \omega}{ ( 2 \pi) ^{d+1 } } \frac{1}{q^2 - a  Log | \omega_n |  }  \nonumber \\
   & \times&    \frac{1}{ ( i \omega_n  + \alpha ^\prime v_F q cos \theta ) ( i \omega_n - v_F q cos \theta ) }  \ , \nonumber \\
     & = &  \frac{ \bar g } {N  Log ( \alpha ^\prime  ) ( 1 - \alpha ^\prime  )  } \int \frac{ q^2 d q d\omega}{ (2 \pi )^3 }  \frac { 1} { - i \omega_n  ( q^2 + a Log |\omega_n |  )}  \times \nonumber \\
     && \left ( \frac{ -1}{ v_F q } \right )  Log \left [ \frac{v_F q - x }{ \alpha^\prime v_F q - x } \right ] _{x=-1}^{x=1}  \ , \nonumber \\
     & = & \frac{ {\bar g }  ( - Log  \alpha ) } {  N ( 1 - \alpha ^\prime )  12 ( 2  \pi ) ^2  E_F  }  \nonumber \   , \eea
with $a = 1/ Log (\alpha^\prime) $. Hence the static vertex is small in $1/N$.

 We  next evaluate the dynamical vertex, with  linearized Fermi surfaces
  \vspace{-30 pt}  
  \begin{center}
    $  \Gamma(q,\Omega) \; $= \begin{picture}(70,70)(0,33)
\Text(12,43)[c]{$q,\Omega$}
\Text(63,2)[c]{$k_F,0$}
\Vertex(20,35){1.5}
\Photon(5,35)(20,35){2}{2.5}
\DashArrowLine(20,35)(50,53){1.5}
\ArrowLine(50,17)(20,35)
\ArrowLine(50,53)(65,62)
\DashArrowLine(65,8)(50,17){1.5}
\Photon(50,17)(50,53){2}{4.5}
\Vertex(50,17){1.5}
\Vertex(50,53){1.5}
\end{picture}
\end{center}
\vspace{30 pt } 
\bea \label{eqnan2}
   && \Gamma(q, \Omega_n)  =   \frac{\bar g }{-  N  Log( \alpha ^\prime )} \int \frac{d^d q d \omega}{ ( 2 \pi) ^{d+1 } } \frac{1}{q^2 - a  Log | \omega_n |  }  \nonumber  \\
   & \times&    \frac{1}{ ( i \omega_n  + \alpha ^\prime v_F q cos \theta ) ( i \omega_n   + i \Omega_n - v_F q cos \theta - v_F Q_x ) }  \ , \nonumber \\ 
   & =&  \frac{\bar g }{-  N  Log( \alpha ^\prime ) ( 2 \pi)^3} \frac{i}{ i \Omega_n - v_F Q_x}  \int_{-\Omega_n}^0 d \omega Log \frac{( - Log \alpha^\prime )} { Log |\omega_n |} , \nonumber \\
   & \simeq &  \frac{\bar g }{-  N  Log( \alpha ^\prime ) ( 2 \pi )^3}  \nonumber \\
   & \times &  \frac{i}{ i \Omega_n - v_F Q_x} \left (  \Omega_n Log Log |\Omega_n| - Li ( - \Omega_n )  \right )  \  . \nonumber \eea We see from (\ref{eqnan2}) that  for   $Q_x=0$ the vertex has  a $Log Log  $- singularity. It is not small.

 The same evaluation with the curvature of the Fermi surface gives:
 \bea \label{eqnan3}
   & & \Gamma(q, \Omega_n)  =   \frac{\bar g }{-  N  Log( \alpha^\prime )} \int \frac{d^d q d \omega}{ ( 2 \pi) ^{d+1 } } \frac{1}{q^2 - a  Log | \omega_n |  }  \nonumber  \\
   & \times&    \frac{1}{ ( i \omega_n  - \alpha^\prime v_F q cos \theta -  \alpha^\prime  q_\perp^2/ (2 m )  ) } \times \nonumber \\
    &  &\frac{1}{ ( i \omega_n   + i \Omega_n - v_F q cos \theta - v_F Q_x  - q_\perp^2/ (2 m ) ) }  \ , \nonumber \\
    & = &  \frac{\bar g }{-  N  Log( \alpha ^\prime ) 2  ( 2 \pi )^3} \int_{- \Omega_n }^0  ( - i d \omega ) \times \nonumber \\
    &  &\frac{1}{ ( Q_x v _F  + ( 1 - \alpha^\prime ) Log |\omega_n|/ Log \alpha^\prime  ) }  \
     Log \left ( \frac{- Log \alpha^\prime}{Log |\omega_n | } \right )  \, \nonumber  \\
    & \simeq & \frac{\bar g }{-  N  Log( \alpha ^\prime ) 2  ( 2 \pi )^3} \ \frac{i \Omega_n Log ( -Log |\Omega_n| ) }{( 1 - \alpha^\prime Log |\Omega_n| / Log \alpha^\prime )}  \  . \nonumber \eea   We see now that the curvature regularizes the  vertex both in large N and in the infra-red frequency sector.

    \section{Instabilities beyond the Eliashberg theory}
    \label{app:BKV}
     In this section, we evaluate  diagrams  beyond Eliashberg theory, but that are  potentially dangerous for the static sector of any 
    $q=0$ QCP.  As mentioned in the main text, such singularities were discovered by BKV\cite{BKV} and are typical of the  type of problems coming from the  presence of a finite Fermi surface  in the theory. Precisely we want to evaluate
     \vspace{- 40 pt}
      \begin{center} $ \Pi_{fc}^{(1)} (T) = $
       \begin{picture}(90,70)(0,88)
\Text(8,105)[c]{$\Pi_a^{(1)}$}
\Vertex(20,90){1.5}
\Vertex(60,90){1.5}
\Photon(5,90)(20,90){2}{2}
\Photon(60,90)(75,90){2}{2}
\ArrowArc(40,90)(20,0,30)
\DashArrowArc(40,90)(20,30,150){1.5}
\PhotonArc(40,128)(30,-120,-60){2}{4}
\ArrowArc(40,90)(20,150,180)
\DashArrowArc(40,90)(20,180,0){1.5}
\end{picture} $ + $  \begin{picture}(90,70)(0,88)
\Text(8,105)[c]{$\Pi_b^{(1)}$}
\Vertex(20,90){1.5}
\Vertex(60,90){1.5}
\Photon(5,90)(20,90){2}{2}
\Photon(60,90)(75,90){2}{2}
\DashArrowArc(40,90)(20,0,30){1.5}
\ArrowArc(40,90)(20,30,150)
\PhotonArc(40,128)(30,-120,-60){2}{4}
\DashArrowArc(40,90)(20,150,180){1.5}
\ArrowArc(40,90)(20,180,0)
\end{picture} \end{center}\vspace{30 pt}   We first note that the two diagrams are proportional: $ \Pi_a^{(1)} = \alpha^\prime \ \Pi_b^{(1)}$ and that there is no corresponding vertex insertion at the first order. One can check that, in the low energy regime $ T \leq  E^*$, the two diagrams are not singular, since the average gap between the spinon and electron Fermi surfaces protects it.  Here we want to check the stability in the intermediate regime for $T\geq E^*$. We have $z=3$ in the boson propagator. 
 To understand the source of the problem it is instructive to compare the  following four-field diagrams
 \vspace{ - 30 pt}
 \begin{center}\begin{picture}(90,70)(0,110)
  \Text(40,110)[c]{$g_{F}$}
 \Text(13,110)[c]{$f$}
  \Text(67,110)[c]{$f$}
   \Text(40,137)[c]{$c$}
    \Text(40,83)[c]{$c$}
 \Vertex(20,90){1.5}
\Vertex(60,90){1.5}
\Vertex(20,130){1.5}
\Vertex(60,130){1.5}
\DashArrowLine(20,130)(20,90){1.5}
\ArrowLine(20,90)(60,90)
\DashArrowLine(60,90)(60,130){1.5}
\ArrowLine(60,130)(20,130)
\Photon(10,80)(20,90){2}{2}
\Photon(70,80)(60,90){2}{2}
\Photon(10,140)(20,130){2}{2}
\Photon(70,140)(60,130){2}{2}
\end{picture}   and   \begin{picture}(90,70)(0,110)
\Text(40,130)[c]{$g_4$}
 \Vertex(40,110){2}
\Photon(30,100)(40,110){2}{2}
\Photon(50,120)(40,110){2}{2}
\Photon(30,120)(40,110){2}{2}
\Photon(50,100)(40,110){2}{2}
\end{picture} \end{center}\vspace{30 pt}  where $g_4$ is a mode-mode coupling constant,coming for example from the term $- J_0 n_i n_j / 4 $ in Eqn (\ref{eqn2}). The $g_4$ mode-mode coupling is  standard in a $\phi^4$-theory and provides correction to scaling  extensively studied in, for example\cite{zinn} chapter 42.   If $g_4 \geq 0$ the  $\phi^4$ theory is stable and, close to a QCP, the corrections to scaling follow the law \vspace{-30 pt} \begin{center} $ m_b(T) $ = \begin{picture}(50,60)(0,32)
\Text(8,15)[c]{$b$}
\Photon(5,20)(20,30){2}{2.5}
\Vertex(20,30){1.5}
\Text(40,30)[c]{$g_4$}
\Photon(20,30)(35,20){2}{2.5}
\PhotonArc(20,40)(10,0,360){2}{10}
\LongArrowArc(20,40)(5,60,120)
\Text(30,15)[c]{$b$}
\end{picture}  $ \sim T ^{(d+z-2)/z} $ \end{center} \vspace{20 pt}  One sees that, in a fermionic theory, one can form corrections to scaling from the fermion vertex $g_F$  which leads to our two diagrams $\Pi_a^{(1)}$ and $\Pi_b^{(1)}$.   The difference between $g_F$ and $g_4$ is that the fermion loop is dangerous, and can change the sign of the vertex. According to the value of $z$, it can as well lead to a more relevant term than the standard $\phi^4$ corrections to scaling.

 We turn to the computation of the diagrams. With ${\bar g } =  8 k_F^2 V^2 /\rho_0$, $c= \rho_0 N/( \alpha^\prime v_F)$.
 \bea
 & & \Pi_a^{(1)} (T)  =   N {\bar g}  \rho_0 T \sum_{n,m \neq 0} \int \  \frac{d^d q}{ ( 2 \pi)^d} d \epsilon_k   D_b(q, \Omega_m ) \nonumber \\
 & \times &  \ G_c^2(k,\omega_n) G_f (k, \omega_n) G_f(k+q, \omega_n + \Omega_m)  \  ;  \nonumber  \\
  & = &  N \rho_0 {\bar g} T \sum_{n,m} \int \  \frac{d^d q}{ ( 2 \pi)^d} d \epsilon_k \frac{1}{c |\Omega_m | / q + a q^2} \nonumber \\
   & \times & \frac{1}{ ( i \omega_n + i \Sigma_c( \omega_n) - \epsilon_k + \mu )^2} \ 
   \frac{1}{ ( i \omega_n + i \Sigma_f ( \omega_n ) - \alpha^\prime \epsilon_k - \epsilon_f ) }  \nonumber \\
    & \times & \frac{1}{ ( i \omega_n  + i \Omega_m + i \Sigma_f( \omega_n ) + i \Sigma_f ( \Omega_m)  - \alpha^\prime \epsilon_k - \alpha^\prime v_F q_x - \epsilon_f ) } \  . \nonumber  \eea     We have ${\rm Sgn }[ \Sigma_{f,c}( \omega_n)] = {\rm Sgn } [\omega_n] $.We perform first the integration over $\epsilon_k$, noticing that the  integral is non zero if the poles are in the same half plane. To fix the ideas we take  $\omega_n < 0$ while $ \omega_n + \Omega_m > 0 $ (the result is identical with the other choice).  We then close the contour in the  upper half plane  to catch the pole  coming from the last factor. We obtain, after neglecting the terms  proportional to $\alpha^\prime$ and  neglecting $i \omega_n$ compared to $i \Sigma_{f,c}( \omega_n)$ in the Green's functions:
    \bea
    & &  \Pi_a^{(1)} (T)  = N  \alpha^\prime {\bar g}  \rho_0 T \sum_{m \neq 0, -m\leq n\leq 0 } \int \  \frac{d^d q}{ ( 2 \pi)^d} \frac{1}{c |\Omega_m | / q + a q^2} \nonumber \\
    & \times& \frac{2 i \pi }{( i \Sigma_{f}( \Omega_n) - \alpha^\prime v_F q_x )}   \nonumber \\
    & \times& \frac{1}{( i \Sigma_{f}( \Omega_n) + i \Sigma_f( \omega_n)  - \alpha^\prime v_F q_x - \epsilon_f )^2} \ ; \nonumber \eea
    Keeping only the dependence in $\Omega_n$ in the integrand, we get
    \bea
    & = &   N \alpha^\prime {\bar g}  \rho_0 T \sum_{m} \int \  \frac{d^d q}{ ( 2 \pi)^d} \frac{i \Omega_m}{c |\Omega_m | / q + a q^2} \nonumber \\
    & \times &  \frac{1 }{( i \Sigma_{f}( \Omega_m) - \alpha^\prime v_F q_x )^3} \ ; \eea  The calculation in $D=2$ and $D=3$ differs at this point.
    \subsection{$D=3$} 
     Since there is only one pole in the last factor, the integration over $q_x$  lead to a typical $q_x$ of the  order of $\Sigma_f$ (it is the reason why full fermion lines can be  safely replaced by  bare ones in the Eliashberg theory). One possibility is that the integration over $q_x$ doesn't vanish due to the branch cut in  the boson propagator  $ q_x\sim q_\perp \sim  |\Omega_m |^{1/3} $.
     It is what happens in $D=3$. We find
     \beq
       \Pi_a^{(1)} (T) \simeq - T^{4/3}  \ . \eeq  Note that the  minus sign is of importance.  We argue though, that since we are in the intermediate regime where $T\gg E^*$ ,  the correction in $T^{4/3}$ is irrelevant, just giving an extra UV cut-off for the intermediate regime.  The stability of this regime is thus ensured in $D=3$.
       \subsection{$D=2$}  In $D=2$ the situation is more complex.  The branch cut  at  $ q_x\sim q_\perp \sim  |\Omega_m |^{1/3} $ would lead to  $  \Pi_a^{(1)} (T) \simeq - T $, but there is a stronger singularity first discovered in Ref.\cite{jerome}. Indeed,we suppose $q_y \gg q_x$ and expand $ q= \sqrt{q_x^2+q_y^2} \simeq | q_y| + q_x^2 / (2 | q_y| ) $. We find
       \bea
 & &  \Pi_a^{(1)} (T)  = N \frac{\alpha^\prime {\bar g}  \rho_0}{(2 \pi)^2} T \sum_{m \neq 0} \int_{|q_x|}^{\Lambda}  \ dq_y \frac{i \Omega_m}{c |\Omega_m | } \nonumber \\
 &\times& \int dq_x \  \left (  | q_y| + q_x^2 / (2 | q_y| ) \right ) \frac{1 }{( i \Sigma_{f}( \Omega_m) - \alpha^\prime v_F q_x )^3} \ ; \nonumber \eea the integration over $q_y$ now leads to a logarithmic singularity in $q_x$
 \[ I_{sing}  = \int d q_x \ \frac{q_x^2 \ {\rm Log}  ( \Lambda / |q_x | ) }{ ( i \Sigma_f( \Omega_m  ) - \alpha^\prime v_F q_x )^3 }  \ . \]  $I_{sing}$ is performed by continuation in the upper half plane, if $\Omega_m \leq 0 $ and in the lower half plane if $\Omega_m \geq 0 $ so that to avoid the pole in the Green's function. Changing variables in $q_x= i z$ we get
 \bea
 I_{sing} & = & - \int_0^\Lambda   i d z    {\rm Sgn }( \Omega_m ) 
 \frac{(- i z)^2 \left ( {\rm Log } ( -i z) - {\rm Log } ( i z) \right ) }{ (-i )^3  \left (| \Sigma_f( \Omega_m ) |  + \alpha^\prime v_F z \right )^3 }\ ;  \nonumber   \\
 & =& \frac{- i \pi {\rm Sgn} ( \Omega_m)}{ ( \alpha^\prime v_F)^3} \ {\rm Log} \left ( \frac{\Lambda}{| \Sigma_f ( \Omega_m  ) | } \right )  \  . \nonumber \eea Putting things together we get
 \bea \Pi_a^{(1)} (T)   & = & N  \frac{\alpha^\prime {\bar g}  \rho_0}{(2 \pi)^2} T \sum_{m \neq 0} \frac{i \Omega_m}{c |\Omega_m | } \nonumber \\
 & \times&  \frac{- i \pi {\rm Sgn} ( \Omega_m)}{ ( \alpha^\prime v_F)^3} \ {\rm Log} \left ( \frac{\Lambda}{| \Sigma_f ( \Omega_m  ) | } \right ) \ ; \nonumber \eea since
 $T \sum_{- \Lambda /T }^{\Lambda/T} 1  = Cst $ we find that the T-dependence of the above comes from the $m=0$ term only. Finally
 \beq \Pi_a^{(1)} (T)  =  - \frac{\alpha^\prime {\bar g}  \rho_0}{(2 \pi)^2} \ \frac{ \pi}{ c ( \alpha^\prime v_F)^3}  T {\rm Log} \left ( \frac{ \Lambda}{T} \right )  \ . \eeq    This result puts the intermediate regime in a fragile situation. This term is of negative sign and dominant compare to $E^*$; it has the potential to destabilize the regime. Note that in $D=2$, corrections to scaling coming from the $g_4$ interactions have the same temperature dependence ($ - T logT$), but with a positive sign. Corrections to scaling  hence compete with $\Pi_a^{(1)} $ and tend to stabilize the fixed point. We have reached here the limit of the Eliashberg theory. Whether the intermediate regime is stable or not in $D=2$ depends on whether we have strong enough corrections to scaling of positive sign. This requires, for example, strong enough ferromagnetic short range fluctuations ($ J_0 < 0 $ in Eqn.(\ref{eqn2})).  The stability  of the intermediate regime is a matter of pre-factors between the two terms.

\section{ IL composition rule from the constraint}
\label{app:IL}

 We can recover simply the IL  composition rules by applying the constraint $n_f + n_b = 1$.
 The external current is associated to the conduction electrons  and to the  bosons b.
 \beq {\bf J } = {\bf J}_c + {\bf J}_b \  , \eeq but introducing now the fictitious field ${\bf e}$ and the 
 external field ${\bf E} $ we get
 \bea \label{eqnIL1}
{\bf J}_c  & =&  \sigma_c \ {\bf  E}  \nonumber \\
{\bf J}_b  & = & \sigma_b \ ( {\bf E}  +  {\bf e} ) \nonumber \\
{\bf J}_f &  = & \sigma_f \ {\bf e} \  .  \eea We apply $ {\bf J} _f + {\bf J}_b = 0 $ to (\ref{eqnIL1})  and get 
\beq {\bf e} = \frac{ - \sigma_b } {\sigma_f + \sigma_b }  \ {\bf E }\ , \eeq which leads to

\beq
\sigma = \sigma_c + \frac{ \sigma_f \sigma_b}{\sigma_f + \sigma_b } \ . \eeq

\section{ Evaluation of the polarization and self-energy of the bosons at the QCP}
\label{app:polarization}

 \subsection{Self-energy}
\vspace{- 35 pt}
  \begin{center}
  $\Sigma_b ( \Omega_n ) = $ \begin{picture}(90,70)(0,28)
\Text(10,20)[c]{$b$}
\Text(35,55)[c]{$ a_\mu$}
\Vertex(50,30){1.5}
\Vertex(20,30){1.5}
\GlueArc(35,30)(15,0,180){2}{6.5}
\LongArrowArc(35,30)(10,60,120)
\Photon(20,30)(50,30){2}{4.5}
\Photon(5,30)(20,30){2}{2.5}
\Photon(50,30)(65,30){2}{2.5}
\end{picture}  ; 
\end{center} 
\vspace{20 pt}
   This self-energy captures the effect of the gauge fields on the boson propagator.  This effect is sub-dominant in all regimes but in regimes I where, because of the gap in the particle-hole continuum, the  only source of damping for the bosons are  the gauge fields. We  evaluate $\Sigma_b $ in regime I where $T \leq E^*$ and the particle-hole continuum is gapped. Renaming $c \rightarrow N \pi/ ( 2 m_f v_f) $ and $a \rightarrow 1/ (2 m_f k_F^2)$, we have
   \bea  & & \Sigma_b( \Omega_n)  =    T \sum_n \int \frac{d^d q}{ ( 2 \pi ) ^d } \ \frac{q^2}{ 2 d m_b^2} \ \frac{ 1}{ -i \omega_n   +  a q^2 } \ \nonumber \\
& \times &  \frac{1}{ c  | \omega_n +  \Omega_n | /q  + a q^2 }  \  . \eea Performing the analytical continuation and integrating over the two branch-cuts $\omega _n = 0 $ and $\omega_n + \Omega_n= 0$,  we get for $\Omega_n \geq 0 $
\bea
 & = & \int \frac{d \omega}{4 i \pi } \ {\rm Coth} \left ( \frac{ \omega}{ 2 T} \right ) \ \frac{ d^d q}{ ( 2 \pi ) ^d }  \ \frac{q^2}{ 2 d m_b^2} \ \nonumber \\
  & \times & \left (  \frac{1 } {(-    \omega + a q^2 - i \delta) } \ \frac{ 1}{ ( c ( - i  \omega + \Omega_n  )/q + a q^2 )} \right . \nonumber \\
   & - & \left . \frac{1 } {(- \omega  + a q^2 + i \delta) } \ \frac{ 1}{ ( c ( i \omega  -  \Omega_n ) /q  + a q^2 )}  \right )  ;\nonumber \eea
  \bea
  & = & \int \frac{d \omega}{4 i \pi } \  {\rm Coth} \left ( \frac{ \omega}{ 2 T}\right )  \ \frac{ d^d q}{ ( 2 \pi ) ^d }   \ \frac{q^2}{ 2 d m_b^2} \ ( i \pi ) \delta ( - \omega + a q^2)  \nonumber \\
  & \times &  \frac{2 a  q^2 }{ a^2 q^4  - ( i c \omega/q -c\Omega_n /q)^2 }  \nonumber \\
  &\simeq &\frac{-1}{2} \  \int \frac{ d^d q}{ ( 2 \pi ) ^d }  \ \frac{q^2}{ 2 d m_b^2 }  \ {\rm Coth} \left ( \frac{ a q^2 }{2 T}\right )   \ \frac{2 a  q^2 }{  ( i c a q- c\Omega_n/q ) ^2 }  \nonumber \  . \eea Since this integral is dominated by large q, we get
  \bea \label{damp2} \Sigma_b (\Omega_n) & \simeq &  \frac{1}{2}  |\Omega_n|^{(d+2)/2} \ \int_{\sqrt{2T}}^{\Lambda} \frac{ d^d x}{ ( 2 \pi ) ^d  }  \ \frac{a }{ 2 d m_b^2}  \ \frac{ x^6 }{( i  x^2 - 1)^2  } \nonumber \\
  & =& ( A  + i B) \ |\Omega_n|^{(d+2)/2} \  , \eea
  \[ \mbox{where} \hspace{0.5 cm} \left \{ \begin{array}{l}
   A = Re \left [  \int_0^\Lambda \frac{d^d x}{ ( 2 \pi)^d}  \ \frac{a}{ 4 d m_b^2}  \frac{ x^6 }{( i  x^2 - 1)^2  }  \right ]  \\
  B=  Im \left [  \int_0^\Lambda \frac{d^d x}{ ( 2 \pi)^d}  \ \frac{a}{ 4 d m_b^2}  \frac{ x^6 }{( i  x^2 - 1)^2  }  \right ]  \end{array} \right .   \] in the limit where $T \rightarrow 0 $. We use in the next section the short-hand notation:
  \beq \label{damp1}
  \Sigma_b (\Omega_n)  = f_0^\alpha \  |\Omega_n |^\alpha \hspace{0.5 cm}
    \mbox{with}  \hspace{0.5 cm}  \begin{array}{l}
  \alpha = (d+ 2)/2 ; \\
  f_0^\alpha = A+ i B \end{array} \ . \eeq

  \subsection{Polarization}
   In the following, we evaluate the polarization bubble of the bosons in the three possible regimes at the QCP.
  \vspace{-30 pt}
  \begin{center}
 \begin{picture}(90,70)(80,85)
\Text(90,100)[c]{${\bf A}_\mu$}
\Text(140,100)[c]{${\bf A}_\nu$}
\Text(115,115)[c]{$b$}
\Text(115,65)[c]{$b$}
\LongArrow(86,85)(92,85)
\LongArrow(134,85)(140,85)
\Vertex(100,90){1.5}
\Vertex(130,90){1.5}
\ZigZag(85,90)(100,90){2}{2.5}
\ZigZag(130,90)(145,90){2}{2.5}
\PhotonArc(115,90)(15,0,180){2}{6}
\PhotonArc(115,90)(15,180,0){2}{6}
\LongArrowArc(115,90)(10,60,120)
\end{picture} \vspace{-20 pt}  $ = \;  \Pi_b ( q , i \Omega_n ) $
\end{center}
\vspace{ 50 pt }
The generic form of the bosonic polarization is 
\bea
& & \Pi_b(Q, i \Omega_n)   =  T\sum_{\omega_n}   \int \ \frac{d^d q}{ ( 2 \pi ) ^d } \times \nonumber \\
& &  \frac{q^2 v_b^2}{d} \  D_b ( q, i \omega_n ) D_b( q+Q, i \omega_n + i \Omega_n )  \  , \eea with $v_b$ the  vertex defined in (\ref{eqn42}).
\subsection{ $T=0$: form used in the gauge propagator}
 We evaluate here the bosonic polarization contributing the the gauge fields propagator. Here $v_b = 1/m_b$.
  At $T=0$ only   undamped holons contribute to the damping of the polarization
  \bea &  & \Pi_b(Q, i \Omega_n ) = \int d \omega \int \ \frac{d^d q}{ ( 2 \pi ) ^d } \ \frac{q^2}{ 2 d m_b^2} \ \frac{1}{ -   i \omega + a q^2 } \nonumber \\
  & \times &  \frac{1}{ -  (   i \omega  + i \Omega_n ) + a q^2 + q Q Cos \theta / m_b } ;  \nonumber \eea with $a = 1/(2 m_b )$
  \bea & = & \frac{ 2 i \pi}{- i } \int \ \frac{\Omega_{d-1}  q^d d q d Cos \theta}{ ( 2 \pi )^d } \ \frac{q^2}{ 2 d m_b^2}  \frac{- \Omega_n + i   q Q Cos \theta / m_b }{ \Omega_n^2 + (  q Q Cos \theta /m_b )^2 }  ; \nonumber \\
  &  & \mbox{ where} \ \Omega_{d-1} \mbox{ is the solid angle of dimension d-1}  \nonumber \\
  & = & \int \frac{\Omega_{d-1} q^{d-1} d q}{( 2 \pi )^{d-1} } \ \frac{q^2}{ 2 d m_b}  \ \frac{  \pi |\Omega_n |}{ Q} \  . \nonumber \eea

\subsection{$T\neq 0 $: form used in transport}
 \subsubsection{  Regime I:  there is a gap between the two Fermi surfaces} 
 The form of the boson propagator is given by (\ref{self1})  with $\Pi_{fc}$ given by  the frequency dependence of (\ref{eqn14}) and $\Sigma_b$ given by (\ref{damp1}).  We see that the particle hole  contribution   $\Pi_{fc}$ to the polarization is undamped in this regime; the only source of damping $\Sigma_b$ comes from the gauge fields.
  For the Kubo formulae, we evaluate the polarization at $q \rightarrow 0 $, we need only to retain the damping part of the polarization $\Sigma_b$. The boson polarization then writes ( with $c = 1$ and $a= 1/(2 m_b)$, $v_b= 1/m_b$ ) 
\bea  & &  \Pi_b( 0, i \Omega )    =    T \sum_n \int  \ \frac{d^d q}{ ( 2 \pi ) ^d } \ 
\frac{q^2}{ 2 d m_b^2}  \frac{ 1}{ f_0^\alpha |\omega_n|^\alpha  +  a q^2 } \ \nonumber \\
& \times &  \frac{1}{ f_0^\alpha |\omega_n + \Omega_n |^\alpha  + a q^2 }  ; \nonumber \\
& & \mbox{ considering the two branch cuts at } \  \omega_n = 0 \nonumber \\
& &  \mbox{ and }  \ \omega_n + \Omega_n = 0 \; \mbox{we get}  , \nonumber 
\eea
 \bea & = & \int \frac{d \omega}{4 i \pi } \ {\rm Coth} \left ( \frac{\omega }{2 T} \right )  \ \frac{ d^d q}{ ( 2 \pi ) ^d }  \ \frac{q^2}{ 2 d m_b^2}  \nonumber \\
  & \times & \left (  \frac{1 } { f_0^\alpha (-i  \omega)^\alpha + a q^2) } \ \frac{ 1}{ ( f_0^\alpha ( - i \omega + \Omega_n )^\alpha + a q^2 )} \right . \nonumber \\
  & - &  \frac{1 } { f_0^\alpha (i  \omega)^\alpha + a q^2) } \ \frac{ 1}{ ( f_0^\alpha ( - i \omega + \Omega_n )^\alpha + a q^2 )}  \nonumber \\
  & + &   \frac{1 } { f_0^\alpha (-i  \omega)^\alpha + a q^2) } \ \frac{ 1}{ ( f_0^\alpha (  i \omega + \Omega_n )^\alpha + a q^2 )} \nonumber \\
  & - & \left .  \frac{1 } { f_0^\alpha (i  \omega)^\alpha + a q^2) } \ \frac{ 1}{ ( f_0^\alpha (  i \omega + \Omega_n )^\alpha + a q^2 )}   \right )  ;\nonumber \eea
  \bea
  & = & \int \frac{d \omega}{4 i \pi } \ {\rm Coth} \left ( \frac{ \omega}{2 T} \right )  \ \frac{ d^d q}{ ( 2 \pi ) ^d }   \ \frac{q^2}{ 2 d m_b^2}   \nonumber \\
  & \times &  \left ( \frac{1}{ f_0^\alpha ( - i \omega)^\alpha + a q^2 } - \frac{1}{ f_0^\alpha (  i \omega)^\alpha + a q^2 }  \right )   \nonumber \\
    & \times &  \left ( \frac{1}{ f_0^\alpha ( - i \omega + \Omega_n)^\alpha + a q^2 } + \frac{1}{ f_0^\alpha (  i \omega + \Omega_n)^\alpha + a q^2 }  \right )   \nonumber \\
  \  . \nonumber 
\eea  We expand the second factor in $\Omega_n$ and take the part proportional to $|\Omega_n|$ ( since the constant part cancels with the tadpole diagram). We get
\bea  & & \Pi_b( 0, i \Omega_n )     \simeq   \int \frac{d \omega}{4 i \pi } \ {\rm Coth} \left ( \frac{\omega}{2 T} \right )  \ \frac{ d^d q}{ ( 2 \pi ) ^d }   \ \frac{q^2}{ 2 d m_b^2}  \ \alpha |\Omega_n | \\
& \times & \frac{ 2 i Sin[\alpha \pi/2 ] \ f_0^\alpha \omega^\alpha }{ \left (  (f_0^\alpha \omega^\alpha  Cos[ \alpha \pi /2] + a q^2 )^2 + (Sin [\alpha \pi /2  ] f_0^\alpha \omega^\alpha )^2 \right )^3} \nonumber \\
& \times& \left [  - 2 \omega^{\alpha -1}  Cos[ ( \alpha -1) \pi/2]  \left (  ( f_0^\alpha \omega^\alpha  Cos[\alpha \pi/2] + a q^2 ) ^2 \right . \right . \nonumber \\
&  - &  \left . ( Sin[ \alpha \pi/2 ]  f_0^\alpha \omega^\alpha )^2 \right )  \nonumber \\
& - &  2 i \omega^{\alpha -1 }  Sin[ (\alpha -1 ) \pi / 2 ]  \left ( - i Sin [ \alpha \pi/2 ]  f_0^\alpha \omega^\alpha \right ) \nonumber \\
& \times & \left . \left (  f_0^\alpha \omega^\alpha  Cos[\alpha  \pi /2 ] + a q^2 \right ) \right ]   \nonumber \  . 
\eea 
From the above formulae and using $\alpha = ( d + 2 ) /2 $, we extract $\tau_b $ in all dimensions to get
\beq \begin{array}{l}
\tau_b \sim -log T \;  \mbox{ in } \; D=2 \\
\tau_b \sim T^{5/4} \; \mbox{ in } \; D=3 \end{array} \ . \eeq

\subsubsection{Regime I$^\prime$:  the two Fermi surfaces intersect each other}
We start now from (\ref{eqn16})  (where we have re-named $\rho_0 c/ ( \alpha^\prime v_F q_0 ) \rightarrow c$
 and $\rho_0 a \rightarrow a$, $v_b= 1/m_b$). We take $ \Omega_n \geq 0 $.
\bea  & &  \Pi_b( 0, i \Omega_n )    =    T \sum_n \int \frac{d^d q}{ ( 2 \pi ) ^d } \ 
\frac{q^2}{ 2 d m_b^2} \ \frac{ 1}{ c | \omega_n  | +  a q^2 } \ \nonumber \\
& \times &  \frac{1}{ c  | \omega_n +  \Omega_n |  + a q^2 }  ; \nonumber\\
& & \mbox{ considering the two branch cuts at } \  \omega_n = 0 \nonumber \\
& &  \mbox{ and }  \ \omega_n + \Omega_n = 0 \  \mbox{we have} \  , \nonumber \eea
\bea
 & = & \int \frac{d \omega}{4 i \pi } \ {\rm Coth} \left ( \frac{ \omega }{2 T} \right )  \ 
\frac{ d^d q}{ ( 2 \pi ) ^d }  \ \frac{q^2}{ 2 d m_b^2} \ \nonumber \\
  & \times & \left (  \frac{1 } {(- c i  \omega + a q^2) } \ \frac{ 1}{ ( -c ( i  \omega -  \Omega_n  ) + a q^2 )} \right . \nonumber \\
  & - &  \frac{1 } {( c i \omega  + a q^2) } \ \frac{ 1}{ ( -c ( i  \omega - \Omega_n )  + a q^2 )}  \nonumber \\
  & + &  \frac{1 } {(- c  i \omega  + a q^2) } \ \frac{ 1}{ ( c (i \omega + \Omega_n )  + a q^2 )}  \nonumber \\
  & - & \left . \frac{1 } {( c  i \omega  + a q^2) } \ \frac{ 1}{ ( c ( i \omega  +  \Omega_n )   + a q^2 )}  \right )  ;\nonumber \eea
  \bea
  & = & \int \frac{d \omega}{4 i \pi } \ {\rm Coth} \left ( \frac{\omega}{2 T} \right ) \ \frac{ d^d q}{ ( 2 \pi ) ^d }  
 \ \frac{q^2}{ 2 d m_b^2} \ \frac{  2 i c \omega} { c^2 \omega^2   + a ^2 q^4 } \nonumber \\
  & \times &  \frac{2 c  \Omega_n + a  q^2 }{  c^2 \omega^2 +( c \Omega_n+  a q^2 ) ^2  }  ; \nonumber \eea
   Since the bosonic  mass cancel out with the tadpole diagram (see Appendix \ref{app:WI} or \ref{app:masses}),  we have to extract
    the part proportional to   $\Omega_n $,   to get 
 \[ \simeq   \int \frac{d \omega}{ \pi } \ {\rm Coth} \left ( \frac{ \omega}{2 T} \right )  \ \frac{ d^d q}{ ( 2 \pi ) ^d } 
\ \frac{q^2}{ 2 d m_b^2} \ \frac{ c^2 \omega \Omega_n}{ ( c^2 \omega^2   + a ^2 q^4  ) ^2 }  \  . \]
From the above formulae, we extract $\tau_b $ in all dimensions to get
\beq \begin{array}{l}
\tau_b \sim -log T \;  \mbox{ in } \; D=2 \\
\tau_b \sim T^{1/2} \; \mbox{ in } \; D=3 \end{array} \ . \eeq

\subsubsection{Regime II  both cases}
 In this regime we start with Eqn.(\ref{eqn15}) for the boson propagator, showing $ z=3$ (with the renaming $ \rho_0/ ( \alpha^\prime v_F) c \rightarrow c$ and $ \rho_0 a \rightarrow a $ ) . We take as well $\Omega_n \geq 0$. Here $v_b$ is a bot more comple since we have to expand $|\omega_n|/ (| q + a| ) $ in the first order in  the vector field $a$ to find $v_b$. We get
 \beq
 v_b = 1/m_b - c \omega_n / q^3  \  .\eeq Since  in the integral below 
  \bea  & &  \Pi_b( 0, i \Omega_n )    =    T \sum_n \int \frac{d^d q}{ ( 2 \pi ) ^d } \ 
\frac{q^2 v_b^2}{ 2 d} \ \frac{ 1}{ c  | \omega_n  |/ q +  a q^2 } \ \nonumber \\
& \times &  \frac{1}{c  | \omega_n  + \Omega_n |/ q  + a q^2 }  ; \nonumber \\
& & \mbox{ considering the two branch cuts at } \  \omega_n = 0 \nonumber \\
& &  \mbox{ and }  \ \omega_n + \Omega_n = 0 \   , \nonumber \eea
\bea
 & = & \int \frac{d \omega}{4 i \pi } \ {\rm Coth } \left ( \frac { \omega}{2 T} \right ) \ \frac{ d^d q}{ ( 2 \pi ) ^d }  \ \frac{q^2 v_b^2}{ 2 d} \ \nonumber \\
  & \times & \left (  \frac{1 } {(-  c i  \omega /q+ a q^2) } \ \frac{ 1}{ ( -c( i  \omega - \Omega_n ) /q + a q^2 )} \right . \nonumber \\
  & - &  \frac{1 } {(  c i  \omega / q  + a q^2) } \ \frac{ 1}{ (  - c( i  \omega - \Omega_n )  /q   + a q^2 )}  \nonumber \\
  & + &  \frac{1 } {(-  c i  \omega / q  + a q^2) } \ \frac{ 1}{ (   c( i  \omega + \Omega_n ) /q   + a q^2 )}  \nonumber \\
  & - & \left . \frac{1 } {( c i  \omega /q  + a q^2) } \ \frac{ 1}{ (   c( i  \omega + \Omega_n ) /q   + a q^2 )}  \right )  ;\nonumber \eea
  \bea
  & = & \int \frac{d \omega}{4 i \pi } \ {\rm Coth } \left( \frac{ \omega}{2 T}  \right ) \ \frac{ d^d q}{ ( 2 \pi ) ^d }  
\ \frac{q^2 v_b^2}{ 2 d} \ \frac{  2  c i  \omega / q} { (c/q)^2 \omega^2   + a ^2 q^4 } \nonumber \\
  & \times &  \frac{2  c  \Omega_n/ q+ a  q^2 }{  (c/q)^2 \omega^2 +( c \Omega_n/q+  a q^2 ) ^2  }  ; \nonumber \\
  &  & \mbox{ and taking the part proportional to } \ \Omega_n,  \ \mbox{ we get } \nonumber \\
    & \simeq &  \int \frac{d \omega}{ \pi } {\rm Coth} \left ( \frac{\omega}{2 T } \right )  
\ \frac{ d^d q}{ ( 2 \pi ) ^d } \ \frac{q^2 v_b^2}{ 2 d} 
\ \frac{( c/q)^2  \omega \Omega_n}{ ( (c/q)^2 \omega^2   + a ^2 q^4  ) ^2 }  \  . \nonumber 
\eea From the above formulae, we extract $\tau_b $ in all dimensions to get
\beq \begin{array}{l}
\tau_b \sim -log T \;  \mbox{ in } \; D=2 \\
\tau_b \sim T^{1/3} \; \mbox{ in } \; D=3 \end{array} \ . \eeq

\section{Ward identities (WI) }
\label{app:WI}
 In this section we derive the WI associated to the gauge invariance  of our theory. When it is not mentioned, the field $\sigma$ has been put to zero  at the QCP. The first point of  interest in to show that  the mass of the gauge fields is zero in the Coulomb phase.  The second point is to show that it is non zero in the Higgs phase. In the case of a purely bosonic gauge theory, this second point is simple and can be found, for example, in the Peshkin-Schroder\cite{peshkin}. In the case of  gauge theory with non relativistic fermions, the point is  more subtle since the gauge fields $a_\mu$, $a_\nu$ are not  only coupled to  the Higgs boson. Gorkov\cite{gorkov}  was the first  to show that  the mass is generated in the Higgs phase, in the case of a supercondcutor. Here we follow Zinn Justin on page 432\cite{zinn}, with a field theoretic derivation of the same result.  The possibility of massless Higgs phase, although non generic, will appear at the end. 

We start from Eqn.(\ref{eqn8})  with the gauge fields both coupled to  the spinons f and holons b.  For simplicity we have put $\sigma=0 $ everywhere since this  parameter is irrelevant at the QCP. 
\bea \label{eqnwi1}
S_0 & = & - \int d^d x  d \tau \sum_\sigma \nonumber \\
& &  f^\dagger_\sigma ( x) \left ( \partial \tau + \frac{(\nabla- i e {\bf a}/c ) ^2}{2 m_f }+ \lambda + E_0  \right ) f_\sigma(x) \nonumber  \\
 & + & b^\dagger(x) \left (\partial \tau + \frac{(\nabla- i e {\bf a}/c ) ^2}{2 m_b }+ \lambda  \right ) b (x) \nonumber \\
 &+& \int d^d x d \tau  (  b(x) f^\dagger(x) c(x)  + h. c. ) + H_c \ . \eea
For each field, we introduce a source-term, like
\beq
\begin{array}{l}
 a_\mu \rightarrow J_\mu ; \\
f \rightarrow {\bar \eta}  ;\\
f^\dagger \rightarrow \eta ; \\
b \rightarrow {\bar J}_b ; \\
b^\dagger \rightarrow J_b \end{array}\eeq so that
\beq S= S_0 + S_{source} \ ,  \mbox{with} \eeq
\[
S_{source} = - \int d^d x d \tau  \left (  a _\mu  J _\mu  + f^\dagger \eta + {\bar \eta} f + {\bar J}_b b  + b^\dagger J_b \right  )  \  . \]  
The part  of the action $S_0$ (\ref{eqnwi1}) is invariant under the gauge transformation (\ref{eqn8b}). Only the source terms $ S_{source}$ are affected by the gauge transformation. Using the linearized form of the U(1) algebra $e^{i \theta} = 1 - i \theta $ we get for the variation of the source terms:                            c                        
\bea \label{eqnwi5}
\delta S_{source}  & = &  - \int d^d x d \tau \left ( \frac{J_\mu}{e}\frac{\partial \theta}{\partial \mu }  + i \theta ( {\bar \eta } f - f^\dagger \eta )  \right . \nonumber \\
& + &  \left . i \theta  ( b {\bar J }_b - b^\dagger J_b ) \right )  \  .  \nonumber  \eea  Now from the change of  variables
\beq\begin{array}{l}
f \rightarrow f ( 1 + i \theta) , \\
b \rightarrow b ( 1 -  i \theta ), \\
a_\mu  \rightarrow  a_\mu  + \frac{\partial \theta}{e \partial x_\mu}  , \end{array} \eeq we check that the whole action $ S=S_0 + S_{source} $
is invariant under the U(1) gauge transformation. Hence $\delta S_{source}= 0 $. Integrating by parts the first term in (\ref{eqnwi5}), one gets one version of the WI:
\beq \label{eqnwi2}
-\partial_\mu J_\mu  + i e ({\bar \eta} f - f^\dagger \eta ) + i e ({\bar J}_b b - b^\dagger J_b ) \ \left ( \vdots \right ) = 0 \ . \eeq  (\ref{eqnwi2})  is applied to any generating functional for correlation functions. We can first use the generating functional of  the source currents $W ( J, \eta ) $.
\beq
-\partial_\mu J_\mu  + i e \left  ( {\bar \eta } \frac{\partial W}{\partial {\bar \eta} } - \frac{\partial W}{\partial \eta}  \eta \right  ) + i e  \left (\frac {\partial W}{\partial {\bar J } _b}  {\bar J}_b  - J_b \frac{\partial W}{\partial J_b} \right )  = 0  \ . \eeq  But we  can also use the generating functional for the vertices $\Gamma$, where $\Gamma $ is a p-leg vertex, which leads to the more useful WI:
\beq \label{eqnwi3}
-\partial_\mu \frac{\partial \Gamma}{\partial a_\mu}  + i e \left ( \frac{\partial \Gamma}{\partial f } f - f^\dagger \frac{\partial \Gamma}{\partial f^\dagger } \right ) + i e  \left ( b \frac{\partial \Gamma}{\partial b } - \frac{\partial \Gamma}{\partial b^\dagger} b^\dagger \right )  = 0  \ . \eeq
  To get a better idea, let's derive the WI  for the two-legs vertex, which is nothing but the total polarization $ \Pi_{\mu \nu } $. We differentiate (\ref{eqnwi3})  with respect to $ a_\nu(y) $ to get
  \bea
 &&   -\partial \mu  \frac{\partial^2 \Gamma}{\partial a_\nu ( y ) \partial a_\mu(x) } \\
 & + &  i e  \left ( \frac{\partial^2 \Gamma }{\partial a_\nu (y)  \partial f(x) } f (x)- f^\dagger (x) \frac{\partial^2 \Gamma}{\partial a_\nu(y) \partial f^\dagger(x) } \right )  \nonumber \\
 & + &  i e   \left ( \frac{\partial^2 \Gamma }{\partial a_\nu (y)  \partial b(x) } b(x)- b^\dagger (x) \frac{\partial^2 \Gamma}{\partial a_\nu(y) \partial b^\dagger(x) } \right )  = 0 \ . \nonumber \eea 
 To get the proper vertices, we Fourier transform and then put the sources to zero. We note here that it is possible to put the sources to zero in the Coulomb phase, but not in the Higgs phase where the  boson acquires  a non zero mean-field value. 
 There are two cases of interest : (i) the Coulomb phase and (ii) the Higgs phase. We  get
 (i) in the Coulomb phase :
 \beq
 - i q^\mu \Gamma^2_{\mu \nu }  ( q, - q )  = 0  \ , \eeq which is re-written,with the notations of the body of this paper
 \beq \label{eqn44}
 -i q^\mu \Pi_{\mu \nu } ( q, - q) = 0 \ . \eeq  From (\ref{eqn44}) we see that, in the  Coulomb phase, the WI ensures that the mass of the transverse gauge field propagator is zero to all orders.   Note that the identity (\ref{eqn44}) constraints only the  mass of the vector fields; since the mass is taken at $q^0= \omega= 0 $, for which the scalar field $a_0$ is  dropping out of (\ref{eqn44}).
 Now (ii), in the Higgs phase, the WI  writes
 \beq
 - i q^\mu \Gamma^2_{\mu \nu} (q, -q)   + i e  \langle b \rangle \left ( \Gamma^2_{b \nu} (q, -q) - \Gamma^2_{b^\dagger \nu} (q, -q )  \right )  = 0 . \eeq We see that the gauge field propagator gets massive as soon as  $  \Gamma^2_{b \nu} (q, -q) - \Gamma^2_{b^\dagger \nu} (q, -q ) \neq 0 $. This phenomenon is nothing but the  Meissner effect for superconductors. Note that it can happen that, for  special  forms of the Fermi surface,
 $  \Gamma^2_{b \nu} (q, -q) - \Gamma^2_{b^\dagger \nu} (q, -q ) = 0 \  . $ We then get the equivalent of  massless superconductivity, for  a U(1) gauge theory.
 
 We can also derive the same kind of WI for the three legs vertex. Let's take for example
  \begin{center}
  $ \Gamma_{a_\mu f^\dagger_{k+ q}f_k} $  = \begin{picture}(50,30) (0,20)
  \Vertex(20,20){1.5}
  \ZigZag(5,20)(20,20){2}{4}
  \ArrowLine(30,30)(20,20)
  \ArrowLine(20,20)(30,10)
  \Text(10,30)[c]{$a_\mu (q)$}
  \Text(40,30)[c]{$f_k$}
  \Text(40,10)[c]{$f^\dagger_{k+q}$}
  \end{picture}
  \end{center} 
  \vspace{ 10 pt}  We differentiate (\ref{eqnwi3}) with respect to $f(y)$ and then to $f^\dagger (z)$ and put the source terms to zero, except in the Higgs phase:
  \bea
   & & -\partial_\mu  \frac{\partial^3 \Gamma}{\partial a_\mu (x) \partial f^\dagger (z) \partial f(y) } \\
   & + & i e \left ( - \frac{\partial^2 \Gamma}{\partial f^\dagger (z) \partial f (x) } \delta_{x y} - \frac{\partial^2 \Gamma}{\partial f (y) \partial f^\dagger (x) } \delta_{x z} \right ) \nonumber \\
   & + &  i e \left ( b (x) \frac{\partial^3 \Gamma}{\partial f^\dagger (z) \partial f(y )  \partial b(x) } \right .\nonumber \\
&-& \left . \frac{\partial^3 \Gamma}{\partial f^\dagger(z) \partial f(y) \partial b^\dagger( x) } b^\dagger (x)  \right ) = 0 \  . \nonumber \eea 
   (i) In the Coulomb phase
   \beq \label{eqnwi4}
   - i q^\mu \Gamma^3_{a_\mu f^\dagger_{k+ q}f_k} + i e \left ( G_f^{-1}(k+ q) - G_f^{-1} (k)  \right )  = 0   \  . \eeq In the Higgs phase (ii) it comes
   \bea
   &&  - i q^\mu \Gamma^3_{a_\mu f^\dagger_{k+ q}f_k} + i e \left ( G_f^{-1}(k+ q) - G_f^{-1} (k)  \right ) \nonumber \\
   & + & i e \langle b \rangle \left ( \frac{\partial^3 \Gamma}{\partial b_q \partial f^\dagger_{k+ q} \partial f_k } - \frac{\partial^3 \Gamma}{\partial b^\dagger_q \partial f^\dagger_k \partial f_{k+q} } \right ) = 0 \nonumber  \  . \eea Quite generically, the p-legs vertex is related to the (p-1)-legs vertex through the  WI.
Note that a relation similar to (\ref{eqnwi4}) can be established for the ferromagnetic QCP, using the translational invariance instead of the gauge invariance;
   one can follow the same steps using the Noether theorem  associated with translation invariance, instead of  the equivalent of it, associated with  U(1) local invariance, that we derived here. 
   
   \section{Direct check of the vanishing of the masses }
   \label{app:masses}
    In this section we directly check that the masses  of the gauge field propagator vanish in the Coulomb phase, at the first order in perturbation theory. 
    \vspace{0.2 cm}   
        \subsection{Fermion part}
         Let's start with the  fermions and check the following cancellation:
    \vspace{- 40 pt}
   \begin{center}
   $ \Pi_f (0,0) $ =  \begin{picture}(90,70)(0,88)
\Text(8,100)[c]{${\bf a}_\mu$}
\Text(59,100)[c]{${\bf a}_\nu$}
\Text(35,115)[c]{$f$}
\Text(35,63)[c]{$f$}
\LongArrow(6,85)(12,85)
\LongArrow(55,85)(60,85)
\Vertex(20,90){1.5}
\Vertex(50,90){1.5}
\ZigZag(5,90)(20,90){2}{2.5}
\ZigZag(50,90)(65,90){2}{2.5}
\DashArrowArcn(35,90)(15,0,180){1.5}
\DashArrowArcn(35,90)(15,180,0){1.5}
\end{picture} $ + $ \begin{picture}(50,50)(0,32)
\Text(8,15)[c]{${\bf a}_\mu$}
\ZigZag(5,20)(20,30){2}{2.5}
\Vertex(20,30){1.5}
\DashArrowArc(20,40)(10,0,360){1.5}
\ZigZag(20,30)(35,20){2}{2.5}
\Text(30,15)[c]{${\bf a}_\nu$}
\end{picture} $= 0$
\end{center}
\vspace{ 30 pt } 
\bea 
\Pi_f (0,0)  & = & \delta_{ij} \frac{T}{2 m_f^2} \sum_{\omega_n} \int \frac{d^d k}{( 2 \pi )^d} \frac{k^2}{d}  \frac{1}{(  i  \omega_n  - \epsilon^0_k )^2 }  \nonumber \\
& + & \frac{T}{2 m_f}  \sum_{\omega_n} \int \frac{d^d k}{( 2 \pi )^d}  \frac{1}{(  i  \omega_n  - \epsilon^0_k ) } \nonumber  \ , \eea  with $\epsilon^0_k = k^2/ ( 2 m_f)  + \epsilon_f $. Relating the first term to the second through the identity
\[ ( - 2 m_f^2) \frac{\partial}{\partial m_f} \frac{1}{( i \omega_n - \epsilon^0_k ) } = \frac{k^2}{ ( i \omega_n  - \epsilon^0_k ) }  \  , \] 
\beq
\mbox{we get } \ \Pi_f(0,0) = \left ( - \frac{1}{d} \frac{\partial} {\partial m_f }  + \frac{1}{ 2 m_f }  \right ) \int \frac{d^d k }{( 2 \pi )^d}  n_F ( \epsilon^0_k)  \  .\eeq 
\[ \mbox{Noticing that} \ 
      \int \frac{d^d k }{( 2 \pi )^d}  n_F ( \epsilon^0_k)  =  \rho_f  
       \sim  m_f^{d/2} \nonumber \ , \] we finally get at the  first order in perturbation theory
            \beq \Pi_f(0,0) = 0  \ .  \eeq
            
  \subsection{ Holon part}  
   The bosonic part follows  the same steps as the fermionic one.
    \vspace{- 40 pt}
   \begin{center}
   $ \Pi_b (0,0) $ =  \begin{picture}(90,70)(0,88)
\Text(8,100)[c]{${\bf a}_\mu$}
\Text(59,100)[c]{${\bf a}_\nu$}
\Text(35,115)[c]{$b$}
\Text(35,63)[c]{$b$}
\LongArrow(6,85)(12,85)
\LongArrow(55,85)(60,85)
\Vertex(20,90){1.5}
\Vertex(50,90){1.5}
\ZigZag(5,90)(20,90){2}{2.5}
\ZigZag(50,90)(65,90){2}{2.5}
\PhotonArc(35,90)(15,0,180){2}{6}
\PhotonArc(35,90)(15,180,0){2}{6}
\LongArrowArc(35,90)(10,60,120)
\end{picture} $ + $ \begin{picture}(50,50)(0,32)
\Text(8,15)[c]{${\bf a}_\mu$}
\ZigZag(5,20)(20,30){2}{2.5}
\Vertex(20,30){1.5}
\ZigZag(20,30)(35,20){2}{2.5}
\PhotonArc(20,40)(10,0,360){2}{10}
\LongArrowArc(20,40)(5,60,120)
\Text(30,15)[c]{${\bf a}_\nu$}
\end{picture} $= 0$
\end{center}
\vspace{ 30 pt } 
\bea \Pi_b (0,0)  & = & \delta_{ij} \frac{T}{2 m_b^2} \sum_{\omega_n} \int \frac{d^d q}{( 2 \pi )^d} \frac{q^2}{d}  \frac{1}{(  i  \omega_n  - q^2/ ( 2 m_b)  )^2 }  \nonumber \\
& + & \frac{T}{2 m_b}  \sum_{\omega_n} \int \frac{d^d q}{( 2 \pi )^d}  \frac{1}{(  i  \omega_n  -  q^2/ ( 2 m_b) ) } \nonumber  \ . \eea Relating the first term to the second term through the identity
\[ ( - 2 m_b^2) \frac{\partial}{\partial m_b} \frac{1}{( i \omega_n - q^2/ ( 2 m_b) ) } = \frac{q^2}{ ( i \omega_n  - q^2/ ( 2 m_b) ) }  \  , \] 
\bea
\mbox{we get } \ \Pi_b(0,0) &  =  & \left ( - \frac{1}{d} \frac{\partial} {\partial m_b }  + \frac{1}{ 2 m_b }  \right ) \nonumber \\
 &\times & \int \frac{d^d q }{( 2 \pi )^d}  n_B( q^2/ ( 2 m_b) )  \  . \nonumber \eea
\bea \mbox{Noticing that} \  & & 
      \int \frac{d^d q }{( 2 \pi )^d}  n_b ( q^2/ ( 2 m_b))  \nonumber \\
      & = &   \Omega_d \frac{m_b^{d/2}}{( 2 \pi)^d}  \int x^{d-1} d x n_B (x^2) \  ,\nonumber \eea
      with $\Omega_d$ the solid angle in d -dimensions,  we finally get at the  first order in perturbation theory
            \beq \Pi_b(0,0) = 0  \ .  \eeq

  \section{Conduction electron's lifetime}
  \label{app:imsigma}         
  
   In this section we give the evaluation of the temperature dependence of the imaginary part of  ${\rm Im } \Sigma_c$. 
   We work in $D=3$, and 
   \beq
   \Sigma_c (T) =  g T \sum_n \int \frac{ d^d q}{ (2 \pi )^d}\ \frac{q}{ c |  \omega_n - \pi T | + a q^3} \ \frac{1}{ i \omega_n - \epsilon^0_{k+q} }  \  , \eeq  where the frequency $\omega_n = ( 2 n + 1) \pi T $. We  close the contour around the two branch cuts for $\omega_n = 0 $ and $\omega_n - \pi T = 0 $  to get
   \bea
    \Sigma_c (T) & = & - g \int \frac{ d \omega} {4 i \pi } \ {\rm Coth} \left ( \frac{ \omega}{ 2 T }  \right )  \ \int \frac{q d^d q} { ( 2 \pi )^d} \nonumber \\
   &  \times & \left  [ \frac{1}{-i c   \omega  + a q^3}  \ \frac{1}{ \omega  + i \pi T  - \alpha v_F q_x }  \right . \nonumber \\
   & - & \left . \frac{1}{i c   \omega  + a q^3}  \ \frac{1}{ \omega  + i \pi T  - \alpha v_F q_x }  \right ] \nonumber \\
   & - & g \int \frac{ d \omega} {4 i \pi } \ {\rm Tanh} \left ( \frac{ \omega}{ 2 T }  \right )  \ \int \frac{q d^d q} { ( 2 \pi )^d} \nonumber \\
 &  \times & \left  [ \frac{1}{i c   \omega  + c \pi T+ a q^3}  \ \frac{1}{ \omega  + i \delta  - \alpha v_F q_x }  \right . \nonumber \\
   & - & \left . \frac{1}{i c   \omega  + c \pi T + a q^3}  \ \frac{1}{ \omega  - i \delta - \alpha v_F q_x }  \right ]  \  ; \nonumber \eea
     \bea
     & = & - g \int \frac{ d \omega} {4 i \pi } \ {\rm Coth } \left ( \frac{ \omega}{ 2 T }  \right )  \ \int \frac{q d^d q} { ( 2 \pi )^d} \nonumber \\
   &  \times & \left  [ \frac{2 i c \omega}{c^2   \omega^2  + a^2 q^6}  \ \frac{1}{ \omega  + i \pi T  - \alpha v_F q_x }  \right ] \nonumber \\
     & - & g \int \frac{ d \omega} {4 i \pi } \ {\rm Tanh} \left ( \frac{ \omega}{ 2 T }  \right )  \ \int \frac{q d^d q} { ( 2 \pi )^d} \nonumber \\
 &  \times & \left  [ \frac{2}{i c   \omega  + c \pi T+ a q^3}  \ \frac{1}{ \omega  + i \delta  - \alpha v_F q_x }  \right ]  \ . \nonumber \eea
 Taking the integration over $q_x$ leads to 
 \bea
 \Sigma_c (T)  & = & \frac{g }{\alpha v_F} \int \frac{ d \omega} {4  \pi } \ {\rm Coth } \left ( \frac{ \omega}{ 2 T }  \right )  \ \int \frac{\Omega_q q^{(d-1)} d q} { ( 2 \pi )^{(d-1)}} \ \frac{i  c \omega}{c^2   \omega^2  + a^2 q^6} \nonumber \\
 & + &  \frac{g }{\alpha v_F} \int \frac{ d \omega} {4  \pi } \ {\rm Tanh } \left ( \frac{ \omega}{ 2 T }  \right )  \ \int \frac{\Omega_q q^{(d-1)} d q} { ( 2 \pi )^{(d-1)}} \ \frac{1}{i c\omega + c \pi T + a q^3} \  ,  \nonumber \eea where $\Omega_d$ is the solid angle of dimension $d$. This integral is dominated by the low energy part of the first term  (the high energy part of the first and second terms cancel out) which  leads to 
 \[
 {\rm Im } \Sigma_c  (T) = \frac{g }{\alpha v_F} \int_{a_{IR}}^T \frac{ d \omega} {2  \pi } \frac{T}{\omega}  \ \int \frac{\Omega_q q^{(d-1)} d q} { ( 2 \pi )^{(d-1)}} \ \frac{  c \omega}{c^2   \omega^2  + a^2 q^6} \ ; \] where $a_{IR}$ is a IR cut-off. Taking $d=3$ and changing variables for $x=q^3/\omega$ we get
 \bea
{\rm Im} \Sigma_c  (T)  & = &  \frac{g}{\alpha v_F c} \int_{a_{IR}}^T d \omega \frac{T}{\omega}   \int_0^\infty \frac{ 4 \pi d x }{3 ( 2 \pi )^3} \ \frac{1}{ 1 + a^2 x^2} \  ; \nonumber \\
& \simeq &   T  {\rm Log } \left ( \frac{T}{a_{IR}} \right )  \  . \eea The question is now to determine the cut-off $a_{IR}$.  Since we work at finite temperature, there are two sources of IR cut off which are  $E^*$ and $m_b (T)$ , with $m_b(T)$ is the temperature dependence of the holon mass. 
\[
a_{IR}= Max[E^*,m_b(T) ] \  . \]
$m_b (T)$ is determined by evaluating the corrections to scaling   
 \vspace{- 40 pt}
   \begin{center}
   $ m_b(T) $ = \begin{picture}(50,60)(0,32)
\Text(8,15)[c]{$b$}
\Photon(5,20)(20,30){2}{2.5}
\Vertex(20,30){1.5}
\Text(40,30)[c]{$g_4$}
\Photon(20,30)(35,20){2}{2.5}
\PhotonArc(20,40)(10,0,360){2}{10}
\LongArrowArc(20,40)(5,60,120)
\Text(30,15)[c]{$b$}
\end{picture}  $ + $ \begin{picture}(90,70)(0,88)
   \Text(8,100)[c]{$g_4$}
\Text(59,100)[c]{$g_4$}
\Text(35,115)[c]{$b$}
\Text(35,63)[c]{$b$}
\Photon(5,90)(20,90){2}{2}
\Photon(50,90)(65,90){2}{2}
\Vertex(20,90){1.5}
\Vertex(50,90){1.5}
\Photon(20,90)(50,90){2}{4}
\PhotonArc(35,90)(15,0,180){2}{6}
\PhotonArc(35,90)(15,180,0){2}{6}
\LongArrowArc(35,90)(10,60,120)
\end{picture}   \end{center} \vspace{-30 pt} 
\begin{center}  $+$ \begin{picture}(90,70)(0,28)
\Text(10,20)[c]{$b$}
\Text(35,55)[c]{$ a_\mu$}
\Vertex(50,30){1.5}
\Vertex(20,30){1.5}
\GlueArc(35,30)(15,0,180){2}{6.5}
\LongArrowArc(35,30)(10,60,120)
\Photon(20,30)(50,30){2}{4.5}
\Photon(5,30)(20,30){2}{2.5}
\Photon(50,30)(65,30){2}{2.5}
\end{picture} $ + $ 
\begin{picture}(100,70)(-5,88)
\Text(8,99)[c]{$b$}
\Text(90,99)[c]{$b$}
\Vertex(20,90){1.5}
\Vertex(80,90){1.5}
\Vertex(40,110){1.5}
\Vertex(60,110){1.5}
\Photon(5,90)(20,90){2}{2}
\Photon(80,90)(95,90){2}{2}
\ArrowArcn(40,90)(20,180,90)
\PhotonArc(50,110)(15,0,180){2}{6}
\DashArrowLine(40,110)(60,110){1.5}
\DashArrowLine(60,70)(40,70){1.5}
\DashArrowArcn(40,90)(20,-90,-180){1.5}
\ArrowArcn(60,90)(20,90,0)
\DashArrowArcn(60,90)(20,0,-90){1.5}
\end{picture} \end{center}\vspace{30 pt}
 where $g_4$ is the coupling constant associated to the $\phi^4$-holon field theory, if  it is there. One 
can check that  the first diagram goes like $T^{(d+ z-2)/z}$ the second  one like $T^{(d+2)/2}$ and the third one like $T^{5/3}$ in $D=3$. Hence in $D=3$
\[m_b(T) \simeq T^{4/3}\  . \]  Note that the lines are full, and that we must use the Eliashberg theory for this check. Hence in this model, in the intermediate energy regime, $m_b(T) \ll E^*$. We get
\beq
{\rm Im} \Sigma_c  (T) \simeq T {\rm Log} \left ( \frac{T}{E^*} \right ) \ . \eeq


\vspace{0.5 cm}
  
\newpage 
 \vspace{0.5 cm}

 
 
 
 
 
 
 
 
 
 
 
 
 
 
 

\begin{thebibliography}{99}
\bibitem{note1}  For some compounds like CeCoIn$_5$  the  field excited level is  found at quite low temperatures, less than 100 K, and the assumption of a well formed Kramers doublet can be questioned .
\bibitem{stewart} G. Stewart, Rev. Mod. Phys. {\bf 56}, 755
(1984); {\bf 73}, 797 (2001).
\bibitem{review-piers} P. Coleman {\it et al.}, J. Phys. Cond. Matter
{\bf 13} R723 (2001).
\bibitem{lohneysen} H. v. L\"oneysen {et al.} Rev. Mod. Phys. {\bf 79}, 1015 (2007).
\bibitem{qimiaosteglich}P. Gegenwart {\it et al.} cond-mat/0712.2045.
\bibitem{note2}  By comparison, the notorious linear resistivity in high temperature superconductors holds only for  a bit more than one decade of energy.
\bibitem{saunders} M. Neumann {\it et al.} Science {\bf 317}, 1356 (2007).
\bibitem{RKKY} M. A Ruderman and C. Kittel, Phys. Rev. {\bf 96}, 99 (1954).
\bibitem{doniach} S. Doniach, Physica B, {\bf 91}, 213 (1977).
\bibitem{lacroix} C. Lacroix and M. Cyrot, Phys. Rev. B {\bf 20}, 1969 (1979).
\bibitem{steglich} F. Steglich {\it et al. } Phys. Rev. Lett. {\bf 43}, 1892 (1979).
\bibitem{levin} A. Auerbach and K. levin, Phys.Rev. Lett. {\bf 57}, 877 (1986); Phys. Rev. B {\bf 34}, 3524 (1986).
\bibitem{millis-lee} A. Millis and P.A. Lee,  Phys. Rev. B {\bf 35}, 3394 (1987). 
\bibitem{hertz} J. A. Hertz, Phys. Rev. B {\bf 14}, 1165 (1976).
\bibitem{millis} A. J. Millis, Phys. Rev. B {\bf 48}, 7183 (1993).
\bibitem{moriya} T. Moriya and T. Takimoto, J. Phys. Soc.  Japan {\bf 64}, 960 (1995).
\bibitem{chubAF} Ar. Abanov, A. V Chubukov and J. Schmalian Avd. Phys. {\bf 52}, 119 (2003).
\bibitem{BKV} D. Belitz, T.R. Kirkpatrick and T. Vojta, Phys. Rev. B {\bf 55}, 9452 (1997). 
\bibitem{jerome} J. Rech, C. P\'epin and A. V. Chubukov, Phys. Rev. B {\bf 74}, 195126 (2006).
\bibitem{rosch} A.  Rosch {\it et al.}, Phys. Rev. Lett. {\bf 79},
159 (1997); A. Rosch {\it ibid} {\bf 82}, 4280 (1999).
\bibitem{pines} N. Curro {\it  et al. }, Phys. Rev. B {\bf 70} 235117 (2004).
\bibitem{qimiao} Q. Si {\it et al.} Nature {\bf 413}, 804 (2001); D.R. Grempel and Q. Si,
Phys. Rev. Lett. {\bf 91}, 026401 (2003).
\bibitem{pankov} S. Pankov, S. Florens, A. Georges and G. Kotliar, Phys. rev. B {\bf 69}, 054426 (2004).
\bibitem{ssv} T. Senthil, S. Sachdev and M. Vojta, Phys. Rev. Lett. {\bf 90}, 216403 (2003);
Phys. Rev. B {\bf 69}, 035111 (2004).
\bibitem{schofield} P. Coleman, J. B. Marston, A. J. Schofield, Phys. Rev. B {\bf 72}, 245111 (2005).
\bibitem{roger} M. Roger , Phys. Rev. Lett. {\bf 64},297 (1990).
\bibitem{misguich} G. Misguich, P. Bernu, C. Lhuillier and D. Waldtmann, 
Phys. Rev. Lett. {\bf 81}, 1098 (1998).
\bibitem{us} I. Paul, C. P\'epin and M. Norman, Phys. Rev. Lett. {\bf 98}, 026402 (2007).
\bibitem{note10} At low energies the authors of \cite{ssv} find a bosonic contribution to the conductivity which  varies like $ - 1/ Log T $.  This contribution is actually a correction  to the conduction electron's residual resistivity $ \rho_0 $. Hence $ -  Log T \ll \rho_0^{-1} $ and the range of validity  in temperature is exponentially small : $ T \ll Exp ( - 1/ \rho_0 ) \ll 1$.  The contribution $ -1/ Log T $ they talk about is 
thus completely undetectable experimentally; this regime is dominated by the residual resistivity of the conduction electrons.
A second comment is that the lifetime of the bosons (\ref{damp2})  differs from the one found in Ref.\cite{ssv}; we find their exponent in $D=2$ but not in $D=3$. 
\bibitem{cath} C. P\'epin, Phys. rev. Lett. {\bf 98}, 206401 (2007).
\bibitem{adel} A. Benlagra and C. P\'epin, cond-mat/0709.0354
\bibitem{FFLO} P. Fulde, R. A. Ferrel Phys. Rev. {\bf 135} A550 (1964); 
A. I. Larkin and Y. Ovchinnikov, Sov. Phys. JETP {\bf 20}, 762 (1965).
\bibitem{rice} T. M. Rice, Phys. rev. B {\bf 2},3619 (1970).
\bibitem{deleo} L. De Leo, M. Civelli and G. Kotliar, cond-mat/0702559.
\bibitem{ping} P. Sun and G. Kotliar, Phys. Rev. Lett. {\bf 91},
037209 (2003).
\bibitem{ferrero} M. Ferrero {\it et al.} Phys. Rev. B {\bf 72},205126 (2005).
\bibitem{knecht} C. Knecht {\it et al.} Phys. Rev. B {\bf 72},081103 (2005).
\bibitem{bunemann} J. B\"unemann {\it et al .} J. Phys. Cond. Mat. {\bf 19}, 436206 (2007).
\bibitem{silke} S. Biermann, L. de'Medici and A. Goerges, Phys. Rev. Lett. {\bf95}, 206405 (2005).
\bibitem{demedici} L. de Medici {\it et al .}  Phys. Rev.
Lett. {\bf 95}, 066402 (2005).
\bibitem{sordi} G. Sordi {\it et al.} Phys. Rev. Lett. {\bf 98}, 196403 (2007).
\bibitem{lee-review} P. A. Lee, N. Nagaosa and X-G. Wen, Rev.
Mod. Phys. {\bf 78}, 17-85 (2006).
\bibitem{anderson} P.W. Anderson, Science {\bf 235}, 1196 (1987).
\bibitem{affleck-marston} J. Marston and I Affleck, Phys. Rev. B
\bibitem{leenagaosa} P. A. Lee and N. Nagaosa, Phys. Rev. B {\bf 46}, 5621 (1992).
\bibitem{hsu}  T. C. Hsu, Phys. Rev. B {\bf 41}, 11379 (1990).
\bibitem{burdin} S. Burdin, D. R. Grempel and A. Georges  Phys. Rev. B {\bf 66}, 045111 (2002).
\bibitem{coleman84} P. Coleman, Phys. Rev. B {\bf 29}, 3035 (1984).
\bibitem{read} N. Read and D. M. Newns, J. Phys. C {\bf
16}, 3273 (1983); N. Read, J. Phys. C {\bf 18}, 2651 (1985).
\bibitem{andersonKondo}P. W. Anderson, Science {\bf 235},  1196 (1987).
\bibitem{nozieres}  P. Nozi\`eres, Ann. de Phys. {\bf 10 }, 19 (1985).
\bibitem{peshkin}M. E. Peshkin and D. V. Schroeder, {\it Quantum Field Theory}, Westview Press, chapter 20.
\bibitem{note8} Note that the gauge invariance seems to be broken  in the expansion to the first order in the gauge fields.  It's not the case since one 
must  expand  as well to the first order in $\nabla \theta$  when checking the gauge invariance of the action.
\bibitem{note4} In order to simplify the discussion we have neglected the field $ \sigma$ in this section. In all rigor it should be taken into account.
\bibitem{note9}  The reader might wonder why the coefficient of $q^2$ is not multiplied by N. The reason is that we can a priori decide to maintain the number of particles  constant when going from $N=2$ to large N. Hence the Fermi wave vector is re-scaled  like  $k_F \rightarrow 2 k_F/N $. The coefficient of $q^2$ comes from the UV sector of the theory, and as such acts as an additional coupling constant. This leaves us the latitude to re-scale it the way we want.
\bibitem{aim} B. L. altshuler, L. B. Ioffe and A. J. Millis,  Phys. rev. B {\bf 50}, 14048 (1994).
\bibitem{il} L. B. Ioffe and A. I. Larkin, Phys. Rev. B {\bf 39}, 8988 (1989).
\bibitem{note5} Note that in this formula,  the masses are band masses and not  thermodynamic masses.
\bibitem{mahan}  G.D. Mahan in {\it Many-Particle physics}, Plenum , p. 696.
\bibitem{note6} In the artificial case where the dispersion is purely linear, the polarization is  massless at  $T=0$.
\bibitem{gorkov} L. P. Gorkov, Sov. JETP {\bf 9}, 1364 (1959).
\bibitem{zinn} J. Zinn-Justin, {\it Quantum field theory and critical phenomena}, Oxford Science Publication.
\bibitem{read3}  N. Read, J. Phys. C {\bf 18}, 2651 (1985).
\bibitem{kim} Ki-Seok Kim, Phys. Rev. B 72, 035109 (2005).
\bibitem{reizer}  M. Yu. reizer,  Phys. Rev. B {\bf 39}, 1602 (1989); Phys. Rev. B {\bf 40}, 11571 (1989).


\end{thebibliography}
\end{document}